\PassOptionsToPackage{table}{xcolor}
\documentclass[sigconf]{acmart}

\usepackage{algorithm}
\usepackage{algpseudocode}
\usepackage{hyperref}
\usepackage{caption}
\usepackage{subcaption}
\usepackage{amsmath}
\usepackage{graphicx}
\usepackage{soul}
\usepackage{multirow}
\usepackage[normalem]{ulem}
\usepackage{enumitem}
\usepackage{balance}

\AtBeginDocument{%
  \providecommand\BibTeX{{%
    \normalfont B\kern-0.5em{\scshape i\kern-0.25em b}\kern-0.8em\TeX}}}

\begin{document}

\title{Gotcha! I Know What You are Doing on the FPGA Cloud: Fingerprinting Co-Located Cloud FPGA Accelerators via Measuring Communication Links}
\author{Chongzhou Fang}
\affiliation{%
  \institution{University of California, Davis}
  \city{Davis, California}
  \country{United States}}
\email{czfang@ucdavis.edu}

\author{Ning Miao}
\affiliation{%
  \institution{University of California, Davis}
  \city{Davis, California}
  \country{United States}}
\email{nmiao@ucdavis.edu}

\author{Han Wang}
\affiliation{%
  \institution{Temple University}
  \city{Philadelphia, Pennsylvania}
  \country{United States}}
\email{han.wang.hw@temple.edu}

\author{Jiacheng Zhou}
\affiliation{%
  \institution{University of California, Davis}
  \city{Davis, California}
  \country{United States}}
\email{jczhou@ucdavis.edu}

\author{Tyler Sheaves}
\affiliation{%
  \institution{University of California, Davis}
  \city{Davis, California}
  \country{United States}}
\email{tsheaves@ucdavis.edu}

\author{John M. Emmert}
\affiliation{%
  \institution{University of Cincinnati}
  \city{Cincinnati, Ohio}
  \country{United States}}
\email{emmertj@ucmail.uc.edu}

\author{Avesta Sasan}
\affiliation{%
  \institution{University of California, Davis}
  \city{Davis, California}
  \country{United States}}
\email{asasan@ucdavis.edu}

\author{Houman Homayoun}
\affiliation{%
  \institution{University of California, Davis}
  \city{Davis, California}
  \country{United States}}
\email{hhomayoun@ucdavis.edu}
\begin{abstract}
In recent decades, due to the emerging requirements of computation acceleration, cloud FPGAs have become popular in public clouds. Major cloud service providers, e.g. AWS and Microsoft Azure have provided FPGA computing resources in their infrastructure and have enabled users to design and deploy their own accelerators on these FPGAs. Multi-tenancy FPGAs, where multiple users can share the same FPGA fabric with certain types of isolation to improve resource efficiency, have already been proved feasible. However, this also raises security concerns. Various types of side-channel attacks targeting multi-tenancy FPGAs have been proposed and validated. The awareness of security vulnerabilities in the cloud has motivated cloud providers to take action to enhance the security of their cloud environments.

In FPGA security research papers, researchers always perform attacks under the assumption that attackers successfully co-locate with victims and are aware of the existence of victims on the same FPGA board. However, the way to reach this point, i.e., how attackers secretly obtain information regarding accelerators on the same fabric, is constantly ignored despite the fact that it is non-trivial and important for attackers. In this paper, we present a novel fingerprinting attack to gain the types of co-located FPGA accelerators. We utilize a seemingly non-malicious benchmark accelerator to sniff the communication link and collect performance traces of the FPGA-host communication link. By analyzing these traces, we are able to achieve high classification accuracy for fingerprinting co-located accelerators, which proves that attackers can use our method to perform cloud FPGA accelerator fingerprinting with a high success rate. As far as we know, this is the first paper targeting multi-tenant FPGA accelerator fingerprinting with the communication side-channel.
\end{abstract}





\maketitle

\section{Introduction}
In recent decades, cloud computing has gained great popularity due to its considerable computation power, storage capacity, and pay-as-you-go features. With public cloud services being used, cloud users do not need to set up and maintain their own infrastructure, which greatly reduces their costs. Also, infrastructure-as-a-service (IaaS) public cloud services that are open to public users enable lots of newly emerging applications that require massive computation power, e.g., simulation~\cite{karandikar2018firesim}, deep learning~\cite{carneiro2018performance}, etc. The increasing computation demands incurred by artificial neural networks further motivate the inclusion of hardware accelerators including GPUs~\cite{strom2015scalable}, FPGAs\cite{knodel2018fpgas}, ASICs \cite{chen2016eyeriss}. Among the various hardware accelerators, CPU-FPGAs have become a prevalent heterogeneous architecture to perform computation-intensive workloads due to their programming flexibility, energy efficiency, and high performance. An increasing number of cloud providers have launched commercialized FPGA-based products in the past five years, like AWS~\cite{amazonamazon}, Microsoft Azure~\cite{MS_Azure}, Alibaba Cloud~\cite{alibaba}. To further maximize the utilization of FPGAs, multi-tenancy FPGA infrastructure in the cloud is potentially preferred by the commercial world, which can improve FPGA utilization efficiency by fitting multiple users' designs onto a single FPGA at the same time. It is widely investigated in academia and a promising infrastructure. %However, it poses new security challenges.

However, with notable benefits comes new security threats. There have been research works regarding security attacks on FPGAs, such as bitstream fault injection~\cite{swierczynski2017bitstream}, hardware trojan~\cite{krieg2016malicious}, rowhammer attacks~\cite{weissman2019jackhammer}, etc. In recent years, remote attacks on cloud FPGAs have also emerged. Security problems are especially severe on multi-tenant FPGA clouds, where circuits from multiple users can be placed on the same FPGA. There are attacks such as power side-channel attacks on remote FPGAs~\cite{moini2021remote,provelengios2019characterizing}, fault attacks~\cite{alam2019ram}, etc. There are also other attacks targeting revealing information about cloud infrastructures~\cite{tian2021cloud, tian2020fingerprinting}. These attacks compromise the security of FPGA cloud users and cloud service providers, causing trust issues about FPGA clouds.

As far as we know, all of the existing attacks targeting cloud FPGA user applications demand the knowledge of the co-located victim FPGA circuits, which is non-trivial. Taking the fault attack proposed in~\cite{krautter2022remote} as an example, as a prototype of a remote fault injection attack, it is based on the assumption that the co-located victim FPGA circuit is an AES circuit. Similarly, Moini \textit{et.al}~\cite{moini2021remote} leverages the power side-channel traces to recover the MNIST inputs~\cite{lecun1998mnist} based on the knowledge that the victim FPGA circuit is a binarized neural network (BNN) accelerator. Hence, such prior attacks do not fully present the vulnerabilities and risks in FPGAs cloud, which can lead to the negligence of security challenges and limit corresponding defence solutions. 

% However, most of the previous research works \cite{} simply assume this prerequisite is fulfilled and focus on attack implementation.
In response, this research aims to explore the possibility of fingerprinting victim circuits with passive side-channel information from the communication link, i.e., Peripheral Component Interconnect Express (PCIe) in shared FPGAs. PCIe~\cite{neugebauer2018understanding} is used to connect peripheral devices including FPGAs with host machines, and is open for user interaction. We present that stressing the shared communication link can help reveal the I/O patterns of victim circuits co-locating in the same FPGA board. The deduced knowledge of victim circuits can further enable prior proposed attacks~\cite{krautter2022remote}.

To achieve this, we design a measurement accelerator (in the remainder of this paper, we refer to FPGA circuits deployed by a users as `accelerators') to stress the PCIe and conduct read/write to host memory blocks in the FPGA cloud (Intel DevCloud~\cite{OpenCLGuide} in this work). Then we measure the PCIe bandwidth of our measurement accelerator when different victim accelerators are running on the same FPGA. Once we collect the side-channel traces from PCIe, we leverage machine learning techniques to train a model to classify the victim accelerators. %To receive a comprehensive analysis of PCIe-based fingerprinting results, we include different types of machine learning classification algorithms to build the correlation between the communication link and victim circuit type.
Furthermore, this work looks into the impact of the contention level from benchmarks on fingerprinting success rate. Lastly, we implement a prototype of the fingerprinting attack to infer the co-located victim FPGA circuit in cloud infrastructures for future research.

In summary, the contributions of this work are listed below:
\begin{itemize} 
    \item We present a new attack targeting multi-tenancy FPGA clouds, which can help attackers gain additional knowledge about applications in the FPGA cloud and aid further attack attempts.
    \item We implement a proof of concept attack accelerator as well as its host program, which is able to capture the unique communication fingerprints of co-located accelerators.
    \item Four classification algorithms are included to obtain comprehensively assessment of closed-world fingerprinting success rate, reaching as high as around $90\%$ by random forest. We also evaluate our method in an open-world setting, where the success rate reaches around $80\%$.
    %\item We investigate the fingerprinting success rate under different levels of PCIe contention from benchmark, demonstrating the different behaviours of victim circuits.
    \item By validating the proposed attack method, we reveal a novel security vulnerability in communication links of heterogeneous computing systems and provide insights into possible enhancements of such basic hardware and software components.
\end{itemize}

The remainder of this paper is organized as follows. Basic background knowledge such as cloud FPGA and FPGA security is introduced in Section~\ref{SecBackground}. The threat model containing our assumptions is provided in Section~\ref{SecThreatModel}. Our attack method and its implementation details are shown in Section~\ref{SecMethod}. Section~\ref{SecEval} offers evaluation results. We provide discussion about several defense approaches and future works.  in Section~\ref{SecDiscussion}. Related works are reviewed in Section~\ref{SecRelatedWork}. Finally, we provide a conclusion in Section~\ref{SecConclusion}. 

\section{Background}\label{SecBackground}
\subsection{Cloud FPGA}
Field programmable gate arrays (FPGAs) are integrated circuits that can be programmed after being manufactured. With its great computation power and reprogrammable feature, it is often used to host hardware circuits for custom applications, such as machine learning accelerators. Recently, FPGA resources are starting to be provisioned by cloud service providers. Cloud FPGAs usually operate in two modes: acceleration-as-a-service (AaaS) and FPGA-as-a-service (FaaS)~\cite{dessouky2021sok}. In AaaS mode, FPGAs are pre-configured by the service provider and are offered to users to only accelerate specific computations. On the other hand, the FaaS mode provides users access to the whole FPGA fabric and enables users to program it remotely with greater flexibility. Recently, the concept of multi-tenancy FPGA clouds start to appear~\cite{dessouky2021sok}, which presents a utilization model where a single FPGA in the cloud can be shared by multiple users and can be accessed by these users at the same time.

Cloud FPGAs provide a convenient way for customers to access high-end FPGA resources remotely. Different cloud FPGA providers are offering different types of FPGA resources, e.g. Intel provides users access to Arria 10 FPGAs and Stratix 10 FPGAs on DevCloud~\cite{devcloud}, AWS provides access to Xilinx Virtex UltraScale+ FPGAs with their F1 instances~\cite{AWS}, Alibaba Cloud provides access to Xilinx Kintex UltraScale FPGAs and Arria 10 FPGAs~\cite{alibaba}, etc. %Intel FPGAs are also available on Microsoft Azure\cite{MS_Azure} and Texas Advanced Computing Center (TACC)\cite{TACC}. 
\subsection{Security Problems of Cloud FPGA}
In multi-tenant FPGA clouds, it has been proposed that circuits from multiple users can be placed on the same FPGA, which makes FPGA resource utilization on the cloud more efficient. However, recent research works have shown that cloud FPGAs are vulnerable to various types of side-channel attacks in a multi-tenant setting. Once the security of these FPGA accelerators is compromised, sensitive data or secret keys they are processing can be revealed, which may lead to unwanted data leakage and potentially harm the profits of cloud providers. The following types of attacks are studied most extensively in literature:

\subsubsection{Long-wire Side-Channel Attack}
Long wires, one type of FPGA routing resources that are used to connect configurable logic blocks (CLBs), have been proved to be a source of side-channel information leakage. In~\cite{longwirebit}, the authors find that when a long wire on FPGA is transmitting a logical $1$, the delay of the nearby long wires is shorter than when it is transmitting a logical $0$. Based on this phenomenon, the authors propose to measure the delay of long wires by connecting ring oscillators (ROs) to them. When the target long wire is transmitting a logical $1$, the delay of the nearby long wires will decrease, which causes the frequency of the ROs to increase. By monitoring the frequency change of the ROs in a fixed time interval, the authors successfully recovered $99\%$ of the bits that are being transmitted in the target long wire. Similarly, in~\cite{sidechannelwithoutphysical} the authors recovered the secret key of an AES implementation using the long-wire side-channel attack.

%In\cite{longwirebit}, the authors found that in FPGAs such routing resources, called “long wires”, influences the delay of nearby wires. They found that if a long wire carries a logical 1, the delay of nearby long lines will be slightly lower than when it carries a logical 0. They used such phenomenon to create a new communication channel that are not physically connected. Imaging two adjacent long wires, one carries the secret information (transmitter), and the other long wire is connected to a Ring Oscillator (RO) to measure the delay of the receiver wire. When the transmitter carries a logical 1, the delay of the receiver wire decreases, therefore the frequency of the RO increases. The authors count the number of RO signal transitions during a fixed time interval and recovered over 99\% of the transmitted bits\cite{longwirebit}. Similarly, in\cite{sidechannelwithoutphysical} the authors successfully recovered the secret key of an AES implementation using such long-wire side channel attack.
\subsubsection{Power Side-Channel Attack}
In certain FPGA circuits (e.g. cryptographic circuits), power consumption may be influenced by data being processed in the circuits hence this information may be monitored and used to recover secrets, e.g. cryptographic keys. Normally, deploying power side-channel attacks requires physical access to the FPGA boards in order to assess the system's power usage.  Although direct access to cloud FPGAs is not achievable, Zhao \textit{et.al}~\cite{powersidechannel} propose a power side-channel attack using FPGA as a power monitor. The authors created an RO-based on-chip power monitor and prove that the RO-based FPGA power monitor may be utilized for a power analysis attack on an RSA crypto module on the same FPGA. Furthermore, in~\cite{highspeedpower}, the authors propose a new design for the RO-based power sensor, which can measure the internal voltage in nanosecond scale. They are able to successfully retrieve the secret key of an AES encryption circuit using the power side-channel.

Power side-channel can also be used for accelerator fingerprinting, as shown in~\cite{gobulukoglu2021classifying}. However, in this paper we will show that communication side-channel can be a better option, which has less stringent requirements for attackers and can achieve better performance.
\subsubsection{PCIe Side-Channel Attack}
PCIe contention side-channel has been utilized before to retrieve secret information from CPU-GPU systems~\cite{tan2021invisible}. In~\cite{tian2021cloud}, the authors used PCIe contention to perform an attack on the AWS server. %The Peripheral Component Interconnect Express (PCIe) standard provides a high-speed, serial, point-to-point connection between the components. 
% Although multiple PCIe slots on the motherboard may advertise the same number of lanes and performance, their locations on the PCIe topology graph may vary, 
The authors observe that the difference in locations of PCIe slots in the PCIe topology can result in disparate latency and bandwidth. Based on this, they are able to detect the bandwidth change when different FPGAs in the same sever attempt simultaneous memory accesses to generate PCIe contention and successfully reverse-engineer the locality of different FPGAs in the same AWS server. However, unlike our work, \cite{tian2021cloud} focuses on revealing infrastructure information instead of revealing information about applications on the same FPGA. 

The existing attacks targeting the FPGA cloud presume the knowledge of co-located victim circuit is provided. While in fact this information is nearly impossible to be obtained directly. In response, our work focuses on inferring co-located FPGA-accelerated workloads using PCIe contention side-channel information. Previously proposed attacks will benefit from our attack method.

\subsection{Intel FPGA Cloud Environment}
DevCloud is a cloud platform managed by Intel~\cite{devcloud} to support research and education about FPGAs, GPUs, AI acceleration, etc. In this paper, all the development is done on DevCloud. We choose DevCloud because it provides us access to high-end commercial FPGA devices including Arria 10 and Stratix 10 FPGAs on the cloud, and we can utilize various off-the-shelf toolchains, including high-level-synthesis (HLS)~\cite{coussy2010high}, OpenCL~\cite{devcloudOpenCLGuide} and OneAPI~\cite{oneapi}. %Users from around the world can access the infrastructure via SSH connection and conveniently deploy their applications. 

Accelerators in Intel FPGAs are usually called accelerator functional units (AFUs), which are connected to an interface layer called FPGA interface unit (FIU). In Intel's host-FPGA systems, PCIe serves as the low-level hardware component, and communication is orchestrated by their core cache interface protocol (CCI-P)~\cite{CCIP}. All these low-level protocol details can be agnostic to developers, enabling them to focus on the development of AFUs. Our attack method also does not rely on features of their low-level implementation, and we don't need to hack into these systems managed by FPGA cloud providers.

% An Intel FPGA accelerator package consists of the accelerator functional unit (AFU) and the FPGA interface manager (FIM). AFU is implemented in FPGA logic to provide acceleration for specific applications to improve performance. In FIM, there is a layer called FPGA interface unit (FIU), which acts as a bridge between PCIe, UPI, and AFU-side interfaces like core cache interface protocol (CCI-P)~\cite{CCIP}. The Intel Xeon processor communicates with FUI through PCIe and UPI. UPI stands for Ultra Path Interconnect, it is the interconnection protocol used to connect Intel cores and other IPs. PCIe~\cite{neugebauer2018understanding} is an interface standard for connecting high-speed components. PCIe is widely used in modern computer systems to connect peripheral devices such as graphic cards, WiFi cards, SSDs, etc.  Intel uses CCI-P protocol to connect AFU and FIU. CCI-P provides an abstraction layer that may be built on top of a number of platform interfaces such as PCIe and UPI, allowing the interoperability of CCI-P-compliant AFU across platforms.

\section{Threat Model}\label{SecThreatModel}
In this paper, we investigate the potential of fingerprinting victim circuits in a multi-user FPGA cloud environment~\cite{provelengios2019characterizing}. We follow certain assumptions that are used in previous works~\cite{alam2019ram,provelengios2019characterizing}. Specifically, multiple circuits implemented by different users are placed together on the same FPGA that connects to the same host. How this can be achieved is similar to cloud instance co-location attacks~\cite{nazari2023adversarial,fang2022repttack,fang2023heteroscore}. There are no direct connections or communications between circuits placed by different users. However, the communication links between hosts and connected FPGAs (i.e., PCIe) are shared and communication modules and protocols on top of physical layer are the same. Our work aims to capture the security issue caused by the sharing of the communication link among multiple users.  

The service providers are assumed to be benign and neutral, i.e., they will not attempt to modify user-uploaded circuits. All applications as well as their host programs are executed as provided. Since our attack accelerator will not perform any sensitive operations, it may require special detection mechanisms to defend. We also assume cloud service providers will not terminate our accelerators and host programs. Since the operations proposed in this paper only involve I/O operations that are seemingly normal and no sensitive operations on other users will be triggered, this can be a valid assumption.

We assume attackers and victims have the same privileges in the cloud, i.e., attackers do not have access to more features than victims do. In our settings, the attackers' goal is to obtain information about co-located user applications, which is an important but missing task in existing cloud FPGA side-channel attack works. We consider two different settings: ($1$) closed-world setting, where attackers can limit the range of accelerators running in the cloud; ($2$) open-world setting, where accelerators unknown to the attacker are involved. In both cases, the attacker has access to a server and an FPGA connected in the same way as in a cloud FPGA server.

Benign users are not supposed to be aware of the existence of malicious attackers in the system and hence will not terminate their own accelerators after our attack accelerators are launched. We assume these victim accelerators are constantly running on the FPGA and processing continuous streams of inputs. This assumption is made solely for convenience, since our attack does not require any timing information regarding victim execution life time. These victim accelerators can be either operating on encrypted data or plain original data, but victims will interact with the host through I/O operations. We aim to show that the difference in I/O access patterns can be captured by our proposed attacker accelerator.

In summary, attacks in our target system can be concluded as follows:
\begin{enumerate}
    \item Non-malicious users submit and deploy their accelerators on the cloud, which will keep running for a relatively long period of time;
    \item Attackers submit their malicious accelerators whose goal is to collect performance trace information of communication links and perform classification tasks to determine the exact type of co-located accelerators.
\end{enumerate}
\section{Method and Implementation}\label{SecMethod}
\begin{figure*}[ht!]
    \centering
    \includegraphics[width=\linewidth]{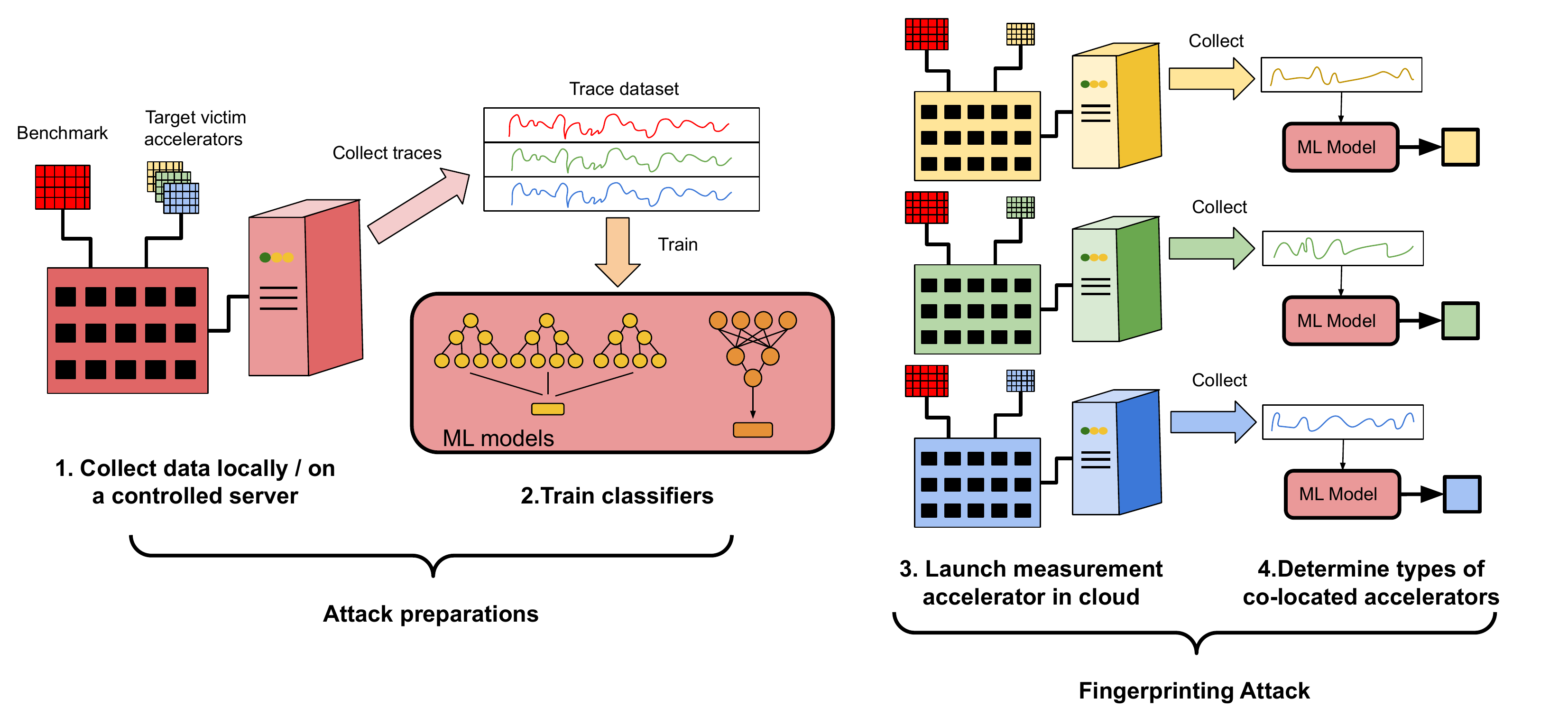}
    \caption{Diagram of our FPGA fingerprinting attack.}
    \label{FigFPGAAttack}
\end{figure*}

In this section, we introduce the design of the proposed fingerprinting attack in FPGA cloud, which consists of attack preparation and online fingerprinting as shown in Figure~\ref{FigFPGAAttack}. The key idea of this work is to capture the execution fingerprints of FPGA circuits by launching a measurement accelerator to measure the bandwidth of communication links and deducing the running victim circuits with machine learning techniques. The whole workflow of our fingerprinting attack consists of several steps:
\begin{enumerate}
    \item Run victim accelerators locally with our proposed measurement circuit to collect data;%on a server with FPGA connected in the same way as the cloud or on a FPGA cloud host to collect data;
    \item Pre-process the collected I/O measurement of possible victim accelerators and train a machine learning-based classifier with the offline collected data set;
    \item Launch the previously used benchmark to the cloud as accelerators and collect I/O measurements;
    \item Pass online I/O traces to the trained classifier to obtain fingerprinting results.
\end{enumerate}

\subsection{Measuring Communication Performance}
In this section, we will introduce the implementation details of the benchmark used to stress the shared communication link and monitor I/O bandwidth. The observation of the benchmark reflects the I/O patterns of co-located victim circuits, which can be further leveraged to reveal the type of victim and used for our proof-of-concept (PoC) fingerprinting attack in the FPGA cloud. We will implement our PoC benchmark accelerator as well as the master host program under OpenCL~\cite{munshi2009opencl} framework. The benchmark consists of two parts: the master host program located in CPU which orchestrates the execution of accelerators, and accelerator circuits in FPGA which stress PCIe communication link. %We will present the details of each part following.

\subsubsection{Host Program Design}
In our PoC benchmark, the host program is responsible for:
\begin{enumerate}
    \item Assign appropriate resources for the operation of accelerator kernels;
    \item Invoke and orchestrate accelerator kernels;
    \item Measure kernel performance using low-level function calls.
\end{enumerate}
There are $3$ design parameters in our host program: \texttt{BUFFER\_NUM}, \texttt{BUFFER\_SIZE} and \texttt{REPEAT\_NUM}. The workflow of our host program is defined as follows. First, the host program will allocate \texttt{BUFFER\_NUM} memory trunks of size \texttt{BUFFER\_SIZE}. Then, these \texttt{BUFFER\_NUM} memory trunks will be accessed by the FPGA accelerator in a pre-defined order. Each of the \texttt{BUFFER\_NUM} memory trunks will be read and written by the accelerator, with traffic passing through the communication link. During the operation to a memory trunk, the time it takes to execute the kernel will be recorded using profiling APIs provided by OpenCL. This information will be further used for calculating the bandwidth of the communication link when all operations to a memory trunk are finished. The operations to a single memory trunk may be repeated for \texttt{REPEAT\_NUM} times and averaged to cancel the effect of noise. Finally, the \texttt{BUFFER\_NUM} measurement of bandwidth will be combined together to form a trace with length \texttt{BUFFER\_NUM}. The pseudo-code for the host program is shown in Figure~\ref{CodeHost}.

\begin{figure}[ht!]
    % \begin{listings}[language=C++]
      \footnotesize
    %   \centering
      \begin{verbatim}
      Allocate buffer[BUFFER_SIZE][1..BUFFER_NUM];
      trace = [];
      for(i = 1..BUFFER_NUM) {
        t_i = 0;
        for(j = 1..REPEAT_NUM) {
            call accelerator and operate on buffer[i];
            t_i += time of kernel execution;
        }
        t_i /= REPEAT_NUM;
        trace.append(1 / t_i);
      }
      return trace;
      \end{verbatim}
      \caption{The pseudo code of our host program.}\label{CodeHost}
    % \end{listings}
    
\end{figure}

\subsubsection{Measurement Accelerator Design}
The measurement accelerator we use in this paper focuses on measuring the I/O bandwidth performance. Similar to previous works that target measuring PCIe performance~\cite{neugebauer2018understanding} or stressing the PCIe connection~\cite{tian2021cloud}, our measurement accelerator implementation follows a similar method and stresses the PCIe communication link via massive read and write communication. There is one design parameter called \texttt{ACCESS\_NUM} that controls how much data is written to the host. The code of our benchmark accelerator implemented as an OpenCL kernel is shown in Figure~\ref{CodeBenchmark}. 

\begin{figure}[ht!]
% \begin{listings}[language=C++]
  \footnotesize
%   \centering
  \begin{verbatim}
  __kernel void mem_kernel(__global int4 *dst) {
      int id = get_global_id(0);
      for(long i = 0; i < ACCESS_NUM; i ++) {
          dst[id] = (int)dst ^ dst[id];
      }
  }
  \end{verbatim}
  \caption{The OpenCL code of our benchmark kernel.}\label{CodeBenchmark}
% \end{listings}
\end{figure}

First, our benchmark accelerator takes in an address pointer (\texttt{dst}). \texttt{dst} is defined as a pointer pointing to pre-allocated host memory. By doing so, we guarantee that our FPGA accelerator will be able to access legally allocated host memory and the generated traffic will pass the FPGA-host communication link. Our kernel then obtains an arbitrarily assigned index to access the host memory space. The exact index is not important in our implementation, and we only use the OpenCL API \texttt{get\_global\_id()} for convenience.

Second, our accelerator enters an execution loop where host memory is accessed multiple times via an array update operation. The same location (\texttt{dst[id]}) in host memory will be updated \texttt{ACCESS\_NUM} times, where \texttt{ACCESS\_NUM} is a design parameter of our benchmark accelerator. The operation listed in Figure~\ref{CodeBenchmark} ensures that a certain amount of data is transferred and the compiler will not optimize out the operation since every time there will be a new value written to the host memory.

\subsection{Data Processing}
In the offline data collection phase (Step $1$ of Figure~\ref{FigFPGAAttack}), the attacker will create a co-location environment and run benchmarks together with potential victim accelerators to collect a performance trace data set. The collected data traces will be normalized and organized in the same data set. Each trace will be labelled according to the types of corresponding victim accelerators.

\begin{figure}[ht!]
    \centering
    \includegraphics[width=\linewidth]{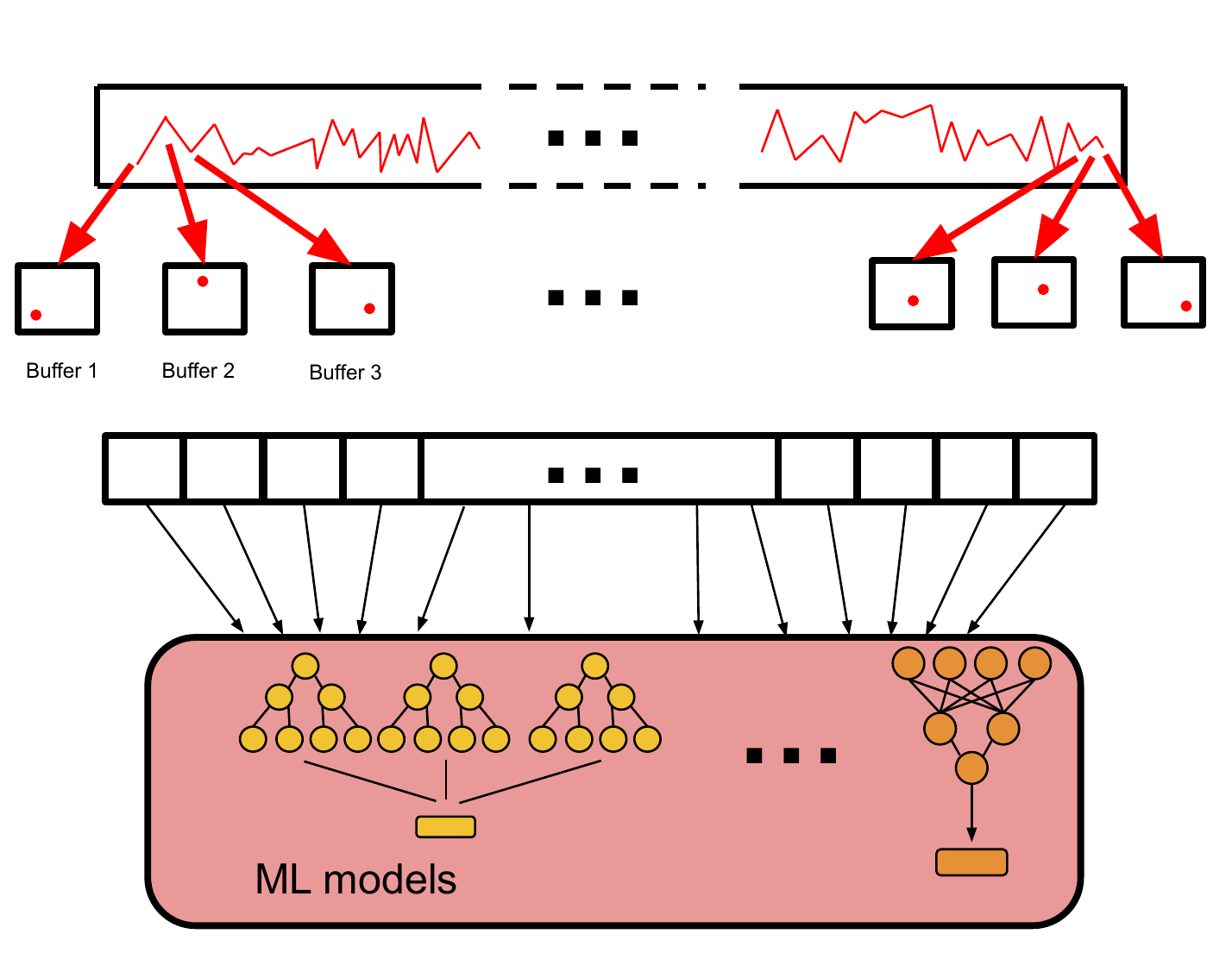}
    \caption{The diagram of our data processing flow.}
    \label{FigProcessing}
\end{figure}

The diagram of our data processing flow is shown in Figure~\ref{FigProcessing}. We can see that each data point within a trace corresponds to the measurement result of kernel execution on an assigned buffer. All the data points will be combined as a feature vector and be fed to machine learning models for further processing.

The resulting data set will consist of all the collected traces, where each row represents one trace. There will be \texttt{BUFFER\_NUM} $+1$ columns in each row, with \texttt{BUFFER\_NUM} entries for collected bandwidth data and $1$ entry for label. The data set will be fed to the machine learning models for training. 

\subsection{Classifiers}
The collected traces are $1$-D vectors with a fixed dimension, since the number of data points are automatically defined by \texttt{BUFFER\_NUM} in the implementation of benchmark circuit. We explore multiple types of machine learning models to assess the potential leakage of the side-channel incurred by the shared communication link. Since we are performing classification tasks and we aim to reduce the costs of attackers by collecting as little data as possible, we select small models that tend to perform well under these scenarios, e.g. Random Forest~\cite{ho1995random} which has been used in fingerprinting tasks~\cite{patwari2022dnn} and also compare the performance of more complex models, e.g. Convolutional Neural Networks. The models we examine in this paper include:
\begin{enumerate}
    \item 1D-Convolution~\cite{kiranyaz20211d}: $1$ convolution layer, followed by a batch normalization layer, a ReLU layer, $3$ layers of fully connected perceptrons~\cite{gallant1990perceptron} and a Softmax layer~\cite{goodfellow2016deep}. 
    \item Multi-layer Perceptron (MLP): $3$ layers of fully-conected perceptrons~\cite{gallant1990perceptron}.
    \item Support Vector Machine (SVM)~\cite{cortes1995support}: classic model that is implemented in popular machine learning libraries~\cite{abadi2016tensorflow,pedregosa2011scikit}.
    \item Random Forest~\cite{ho1995random}: classic model that is implemented in popular machine learning libraries~\cite{abadi2016tensorflow,pedregosa2011scikit}.
\end{enumerate}

For more practical usage in real world, i.e., the open-world scenario, we find that Random Forest can still achieve relatively high accuracy rates even with the existence of unseen accelerator traces. We will demonstrate this in our later evaluation.

\subsection{Implementation of PoC}
The PoC system is built on Devcloud~\cite{devcloud}, using Intel FPGA SDK for OpenCL~\cite{OpenCLGuide}. In the OpenCL toolchain, every OpenCL kernel will be synthesized into customized FPGA circuits. We choose OpenCL because: (1) compared to traditional Verilog RTL design flow, C/C++-based development is faster and sufficient since we don't need to optimize for performance; (2) according to Intel's documentation~\cite{devcloudOpenCLGuide}, as long as different kernels are executed in different OpenCL command queues, they can be executed concurrently which conveniently creates an application co-locating environment that satisfies our need. Besides, we chose to perform our experiments on Devcloud because commercial cloud providers have yet to deploy multi-tenancy FPGAs, despite the possibility of their availability in the future. Nonetheless, our research demonstrates the serious danger posed by the PCIe side-channel when multi-tenancy FPGAs become accessible to users.

In our PoC implementation, victim kernels and accelerator kernels will be defined as two unrelated OpenCL kernels running concurrently on the same FPGA. Victim kernels are configured to run continuously to model victim accelerators that process data streams. They operate on host memory spaces different from those allocated for our benchmark accelerator.

Since the execution environment for offline data collection and online attack is the same, in our PoC implementation the collected data set will be divided to training set and test set, where training set will be used to train the classifier models and the classification results on the test set can emulate the classification accuracy of online attack.
\begin{table*}[!t]
\renewcommand{\arraystretch}{1.3}
% if using array.sty, it might be a good idea to tweak the value of
% \extrarowheight as needed to properly center the text within the cells
\caption{Descriptions of our victim accelerators.}
\label{TabDesAccel}
\centering
% Some packages, such as MDW tools, offer better commands for making tables
% than the plain LaTeX2e tabular which is used here.
\resizebox{\linewidth}{!}{
\begin{tabular}{ccc}
\hline
Name & Code & Function \\ \hline
\rowcolor{lightgray} \texttt{adder} & \texttt{A} & Adder implemented using FPGA. It reads inputs from input buffer, computes results and writes back to a output buffer.\\
\texttt{apply\_watermark} & \texttt{AW} & Image processing circuit. It reads an image from input buffer, adds a watermark andhiheh hack to a output buffer.\\
\rowcolor{lightgray} & & Signal processing circuit. It reads input and coefficient data from input buffer \\
\rowcolor{lightgray} \multirow{-2}{*}{\texttt{fir}} & \multirow{-2}{*}{\texttt{F}} & and performs finite impulse response (FIR) filtering, then writes output back to output buffer.\\
\texttt{matmul} & \texttt{M} & Matrix multiplication circuit. It reads two matrice $A$ and $B$ from input buffer, calculates $AB$ and writes back to output buffer.\\
\rowcolor{lightgray}  & & Convolution accelerator. It reads an image and filter weights from input buffer, \\
\rowcolor{lightgray} \multirow{-2}{*}{\texttt{convolute}} &\multirow{-2}{*}{\texttt{C}} & performs convolution and writes the results back to output buffer.\\
& & This accelerator reads two arrays from input buffer, performs parallel vector addition \\
\multirow{-2}{*}{\texttt{vector\_addition}} & \multirow{-2}{*}{\texttt{V}} & on the two buffers and writes the results back to output buffer.\\
\rowcolor{lightgray} \texttt{noisegen} & \texttt{NG} & An accelerator that generates random traffic between host and FPGA.\\
& & An accelerator employed from Rodinia benchmark~\cite{che2009rodinia}\\
\multirow{-2}{*}{\texttt{hotspot}} & \multirow{-2}{*}{\texttt{HS}} & that performs thermal simulation by iteratively solving differential equations.\\
\hline
\end{tabular}
}
\end{table*}

\section{Evaluation}\label{SecEval}

\subsection{Experiment Settings}
\subsubsection{Hardware Environment}
All our FPGA-related experiments are conducted on Intel DevCloud~\cite{devcloud}. DevCloud allows free SSH access to their servers and FPGA resources from academic users. In our experiments, we select to use nodes with Xeon CPUs and Intel Arria 10 series FPGAs. The environment version is development stack release 1.2.1. To compile our OpenCL kernels, we utilize the tool-chain provided by Intel, which is available on these nodes. Results are all obtained from node named \texttt{s005-n007}.
\subsubsection{I/O Measurement Collection}
The experiment process is controlled by our host program. After the initial setup of OpenCL environments (getting platform information, setting up context, command queues, etc.), we launch a victim kernel which will run continuously during the experiment. Meanwhile we also launch the proposed benchmark to perform multiple measurements on the PCIe communication link and gather data. In each measurement, we initialize a new memory buffer item in host memory and execute benchmark kernel for \texttt{BUFFER\_NUM} times, aggregate acquired data and calculate the average bandwidth as the result of this measurement. For each accelerator, we collect $50$ traces, with \texttt{BUFFER\_NUM} points in each trace (default value $100$). Since victim FPGA circuits and our benchmark circuits (both are synthesized from OpenCL kernel implementation) run concurrently and there is no synchronization step between the two kernels, our experimental setting resembles a multi-tenant FPGA cloud setting.

\subsubsection{Victim Accelerators} In our experiments, we select $8$ different FPGA-accelerated workloads provided by Xilinx Vitis Accelerator Example repository~\cite{xilinx} and FPGA-synthesized GPU workloads from Rodinia benchmark~\cite{che2009rodinia}. We make necessary modifications to deploy them on DevCloud. %The $6$ workloads are: \texttt{adder}, \texttt{apply\_watermark}, \texttt{fir}, \texttt{matmul}, \texttt{convolute}, and \texttt{vector\_addition}.
Detailed description and abbreviation codes are listed in Table~\ref{TabDesAccel}. These accelerators cover different critical areas for FPGA accelerators, including image processing, signal processing, numerical simulation and neural network acceleration thus can serve as representative workloads.
% In our experiments, we target $6$ different accelerators from Xilinx Vitis Accelerator Example repository~\cite{xilinx} and online. We are able to make necessary modifications and deploy them on DevCloud. These accelerators cover different critical areas for FPGA accelerators, including image processing, signal processing and neural network acceleration. The names and functions of these victim accelerators are listed in Table~\ref{TabDesAccel}.

\subsubsection{Classifier Settings}
The collected data will be further analyzed by our learning models. In our experiment, we build several different models based on Python machine learning libraries like Pytorch~\cite{paszke2019pytorch} and Scikit-learn~\cite{pedregosa2011scikit}. The configurations of these models are listed as follows:
\begin{itemize}
    \item Random-forest: \texttt{RandomForestClassifier()} from Scikit-learn library~\cite{pedregosa2011scikit} is used.       
    \item SVM: \texttt{SVC()} classifier from Scikit-learn library~\cite{pedregosa2011scikit} is used.
    \item MLP: built in Pytorch~\cite{paszke2019pytorch}, using learning rate $0.001$, cross-entropy loss and stochastic gradient descent (SGD) optimizer, being trained for $1500$ epochs.
    \item 1D-Convolution: built in Pytorch~\cite{paszke2019pytorch}, using learning rate $0.001$, cross entropy loss and Adam optimizer~\cite{kingma2014adam}, being trained for $1500$ epochs.

\end{itemize}

\subsubsection{Attacking Scenarios}
In this paper, we consider two attacking scenarios, i.e., closed-world and open world scenario. For closed-world testing, we assume all accelerators are known to the attacker, and we consider the fingerprinting problem as an $n$-way classification problem, with $n$ being the number of accelerators in this closed-world. For the more realistic open-world scenario, we consider it as a binary classification problem (since most attackers will have only one specific target for side-channel attacks) and each classifier will be trained to recognize a single accelerator. 

Under both attacking scenarios, we split all collected traces into $7:3$ for training and testing respectively to obtain accuracy data. Additionally, for open-world scenario, we manually remove traces from certain classes in the training set, making these accelerators agnostic to the classifier. The test set will still include traces from these classes to simulate the real-world scenario, where traces from unknown accelerators are collected. %For all of our experiments, training set and test set ratio is $7:3$.

% For closed-world testing, we assume all accelerators are known to the attacker, and we consider the fingerprinting problem as an $n$-way classification problem, with $n$ being the number of accelerators in this closed-world. For the more realistic open-world scenario, we consider it as a binary classification problem (since most attackers will have only one specific target for side-channel attacks) and each classifier will be trained to recognize a single accelerator. We manually remove traces from certain classes in the training set, making these accelerators agnostic to the classifier. The test set will still include traces from these classes to simulate the real-world scenario, where traces from unknown accelerators are collected. For all of our experiments, training set and test set ratio is $7:3$.

% (don't forget to mention that convergence rates are different for differnt networks)

In our experiments, we aim to answer $2$ research questions (RQs):
\begin{enumerate}%[label=\arabic*.]
    \item \textit{RQ1:} Does our measurement circuit capture the communication patterns, and what is the accuracy of fingerprinting? 
    \item \textit{RQ2:} How do the parameter settings of benchmark impact the attacking results?
    % \item How vulnerable are the victim circuits in terms of PCIe-based fingerprinting?
    % \item \textbf{RQ3:} How well do our classification model perform in the classification tasks?
    % \item Is our method scalable?
\end{enumerate}

\subsection{Results}
\subsubsection{RQ $1$:}
\begin{figure}[htbp!]
    \centering
    \begin{subfigure}[t]{0.55\linewidth}
         \centering
         \includegraphics[width=\linewidth]{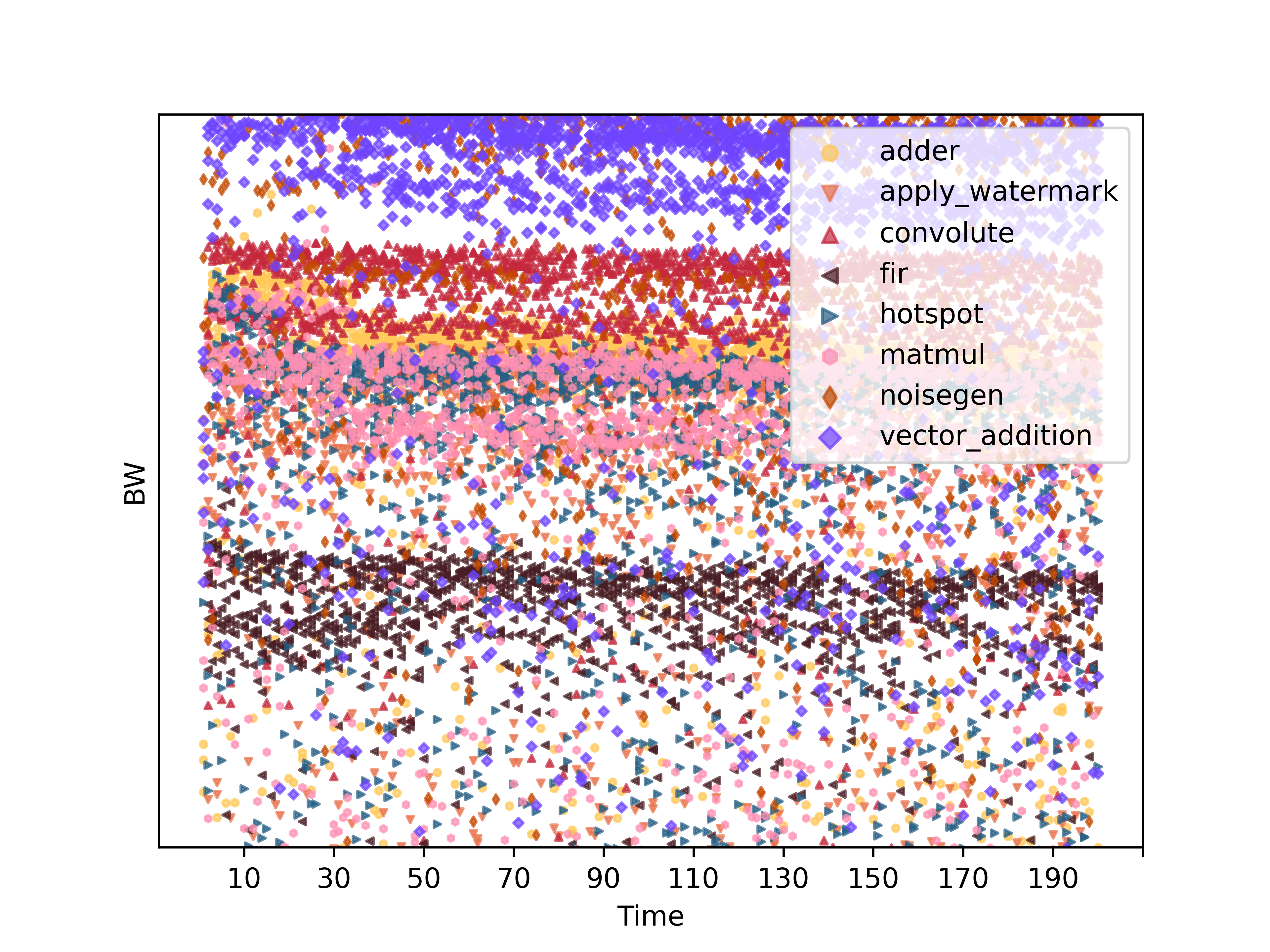}
         \caption{Data collected.}
        %  \label{FigInstNum:b}
     \end{subfigure}
    \begin{subfigure}[t]{0.43\linewidth}
         \centering
         \includegraphics[width=\linewidth]{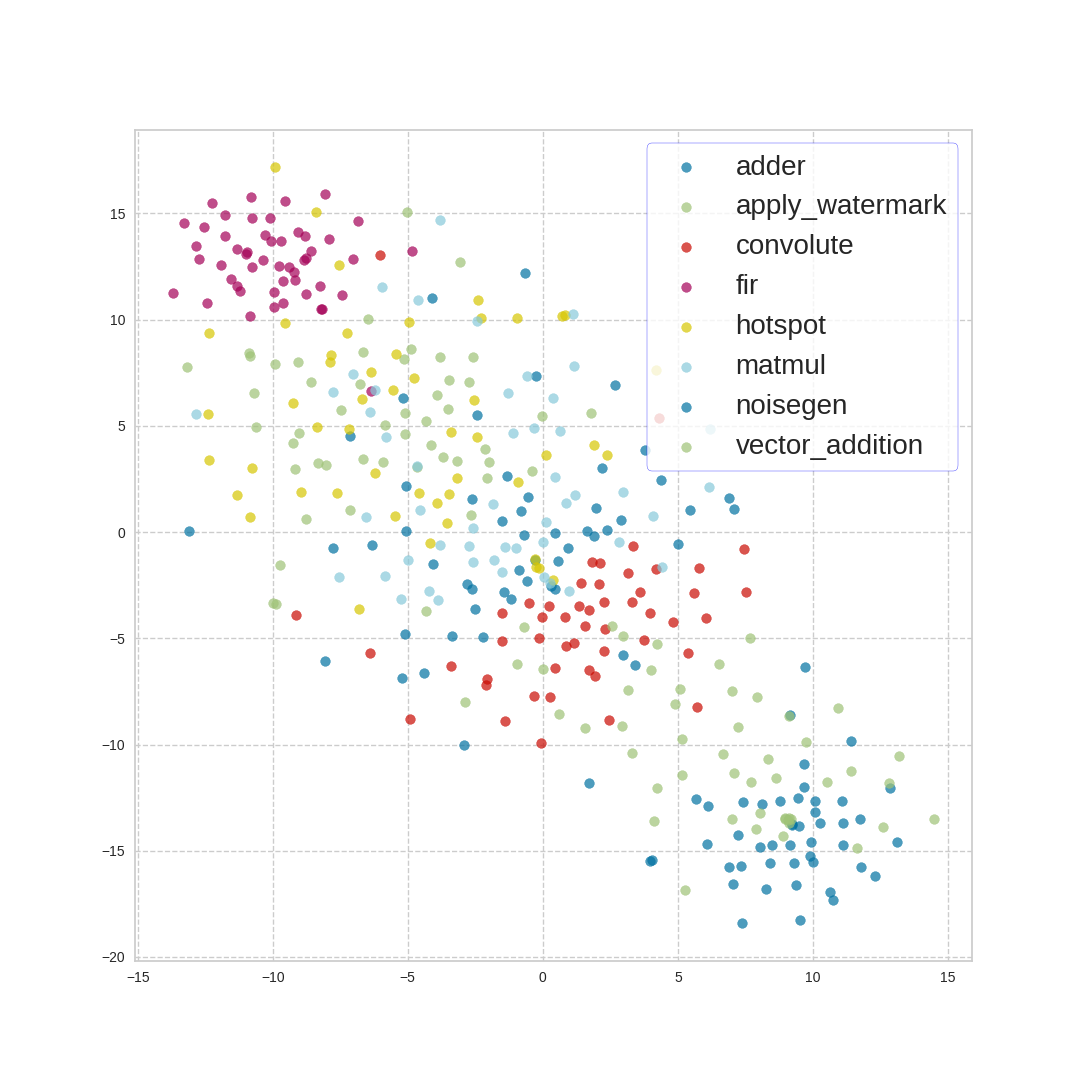}
         \caption{Corresponding t-SNE visualization.}
        %  \label{FigInstNum:b}
     \end{subfigure}
    \caption{Plain data visualization and t-SNE visualization of performance traces collected by our benchmark.}
    \label{FigTrace}
\end{figure}

To answer RQ $1$, we first present the visualization of measured PCIe side-channel traces and fingerprinting accuracy in Figure~\ref{FigTrace} and Table~\ref{TabAccu}, respectively. All data are obtained from the $8$ FPGA-accelerated workloads mentioned above. 

We first present the collected traces with using t-Distributed stochastic neighbour embedding (t-SNE)~\cite{van2008visualizing}, which is a widely used data visualization method. It projects high-dimensional data to a $2$-D plane and can show how the data points can be clustered. In Figure~\ref{FigTrace}, we collect and visualize the bandwidth traces of our benchmark accelerator when it is running concurrently with one of the $8$ victim circuits. In Figure~\ref{FigTrace}~(a), bandwidth data are normalized with minimum and maximum bandwidth values in the data set and range between $0$ and $1$. We can see that traces belonging to different accelerators are separable, which indicates that our benchmark circuit is able to capture the unique communication patterns existing in the execution of the victim accelerators and generate fingerprints for each of them. The bandwidth difference exposes a vulnerability of inferring the co-located victim circuit. T-SNE results in Figure~\ref{FigTrace}~(b) also prove that the data traces are separable.%Following we will leverage the findings to investigate the potential of the leakage and 
% From Figure~\ref{FigTrace}, we can see that there is a human-perceivable layering phenomenon, which indicates that our benchmark circuit is able to capture the communication patterns existing in the execution of the victim accelerators and generate fingerprints for each of them. 

Based on findings in Figure~\ref{FigTrace}, which indicates that these traces contain information that can help differentiate different accelerators, we further consider two fingerprinting attacking scenarios, i.e. closed-world and open-world scenarios. Closed-world fingerprinting aims to classify the types of accelerators within a known accelerator set, while open-world fingerprinting only interests in one sensitive accelerator and classify others as "unrelated".

\textbf{Closed-world.}
\begin{table}[ht]
\begin{minipage}{0.3\linewidth}
\centering
\begin{tabular}{cc}
    \hline
        Model & Test Acc. \\
        \hline
        Random Forest & $\mathbf{88\%}$\\
        SVM & $69\%$\\
        MLP & $55\%$\\
        1D-Convolution & $26\%$\\
        \hline
        %  & Below $70\%$ in realistic settings\\
        % \cite{gobulukoglu2021classifying} & $91\%$ in an un-realistic data set\\
        %  & $97.6\%$ in one case to classify $4$ circuits\\
        % \hline 
    \end{tabular}
    \caption{Accuracy results.}
    \label{TabAccu}
\end{minipage}\hfill
\begin{minipage}{0.55\linewidth}
\centering
\includegraphics[width=.9\linewidth]{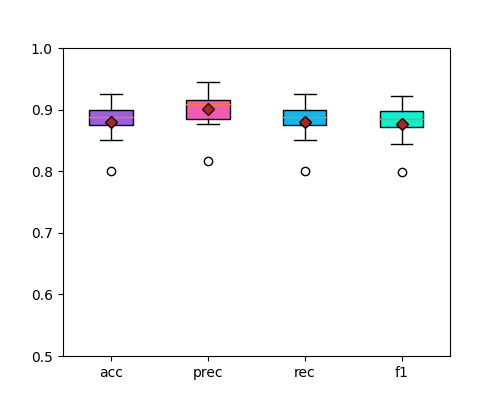}
\captionof{figure}{Cross-validation results of Random Forest model.}
\label{FigCrossValidation}
\end{minipage}
\end{table}

Table~\ref{TabAccu} shows the closed-world classification accuracy performance of our selected models. Among our models, Random Forest achieves the highest classification accuracy, reaching $88\%$ accuracy in this task. The $10$-fold cross-validation results of our Random Forest model is provided in Figure~\ref{FigCrossValidation}. From the distribution of model metrics like accuracy, precision, recall and F1-score we can see that the model, we can see that the model performance is relatively stable. There are similar but different FPGA workload fingerprinting works~\cite{gobulukoglu2021classifying,drewes2023turn}, where the authors utilize power side-channel to perform classification of different cryptographic cores. Compared to their works, we focus on fingerprinting general computing circuits and utilize a different side-channel. Also, our implementation stays at HLS level.
% \begin{table*}[!t]
% % increase table row spacing, adjust to taste
% \renewcommand{\arraystretch}{1.3}
% % if using array.sty, it might be a good idea to tweak the value of
% % \extrarowheight as needed to properly center the text within the cells
% \caption{Metrics of $2$ of our classifier models: SVM and random forest.}
% \label{TabMetrics}
% \centering
% \begin{tabular}{cccccc}
%     \hline
%     Model & Accuracy & Class & Precision & Recall & F-1 Score \\
%     \hline
%     \multirow{6}{*}{SVM} & \multirow{6}{*}{$68.3\%$} & \texttt{adder} & $0.65$ & $0.93$ & $0.76$ \\
%     & & \texttt{apply\_watermark} & $0.56$ & $0.71$ & $0.63$ \\
%     & & \texttt{convolute} & $0.53$ & $0.89$ & $0.67$\\
%     & & \texttt{fir} & $0.91$ & $1.0$ & $0.95$ \\
%     & & \texttt{hotspot} & $0.5$ & $0.21$ & $0.3$ \\
%     & & \texttt{matmul} & $0.62$ & $0.57$ & $0.59$\\
%     & & \texttt{noisegen} & $0.92$ & $0.67$ & $0.77$ \\
%     & & \texttt{vector\_addition} & $0.64$ & $0.58$ & $0.61$\\
%     \hline
%     \multirow{6}{*}{Random Forest} & \multirow{6}{*}{$\mathbf{89.2\%}$} & \texttt{adder} & $0.80$ & $0.92$ & $0.86$ \\
%     & & \texttt{apply\_watermark} & $0.86$ & $0.92$ & $0.89$ \\
%     & & \texttt{convolute} & $1.0$ & $0.88$ & $0.93$\\
%     & & \texttt{fir} & $0.94$ & $1.0$ & $0.97$ \\
%     & & \texttt{hotspot} & $0.58$ & $0.88$ & $0.70$ \\
%     & & \texttt{matmul} & $1.0$ & $0.68$ & $0.81$\\
%     & & \texttt{noisegen} & $0.95$ & $0.95$ & $0.95$ \\
%     & & \texttt{vector\_addition} & $0.94$ & $0.94$ & $0.94$\\
%     \hline
% \end{tabular}
% \end{table*}

\begin{figure*}[htbp!]
    \centering
    \begin{subfigure}[t]{0.48\linewidth}
     \centering
     \includegraphics[width=\linewidth]{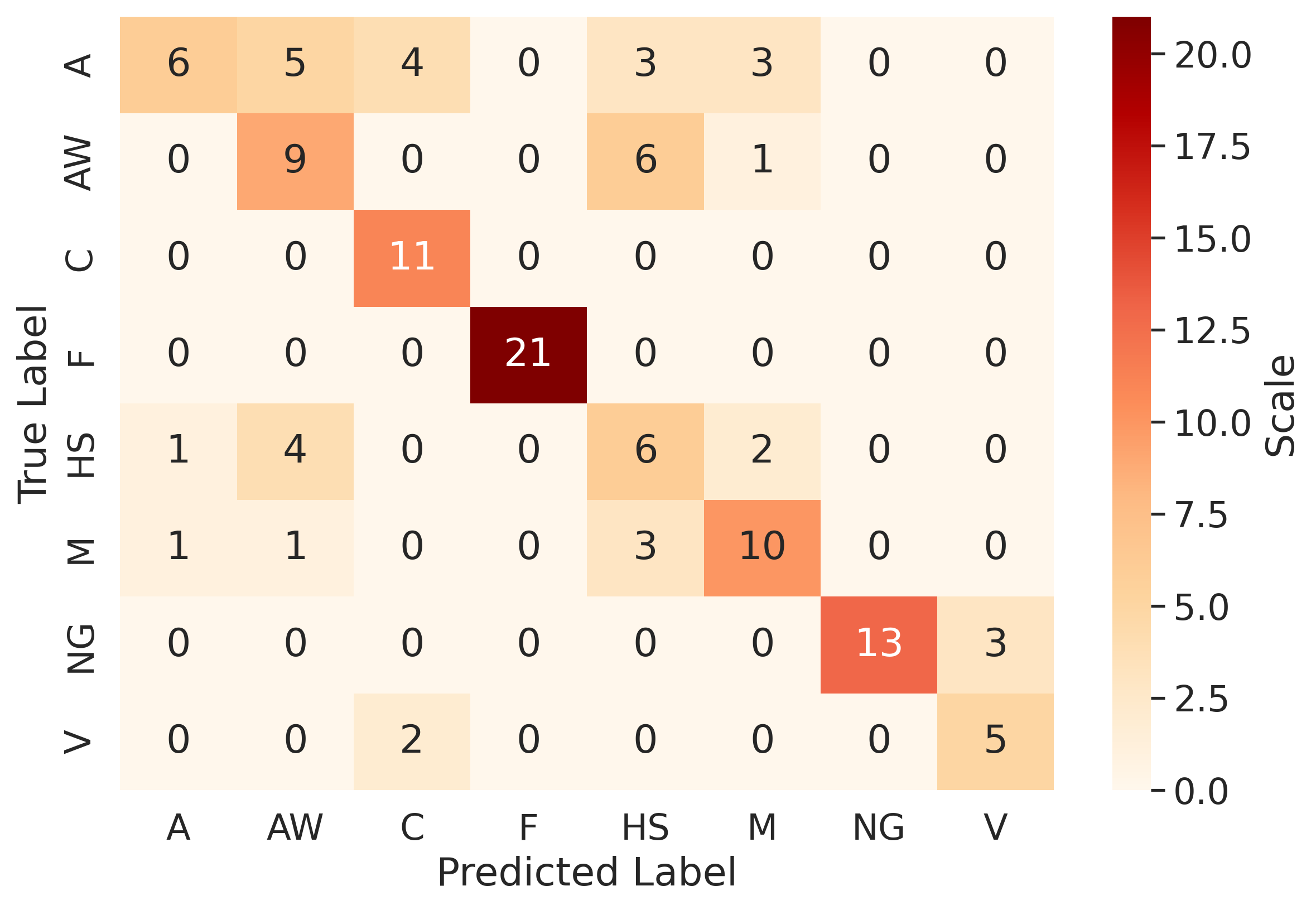}
     \caption{Confusion matrix of SVM classifier.}
 \end{subfigure}
    \begin{subfigure}[t]{0.48\linewidth}
     \centering
     \includegraphics[width=\linewidth]{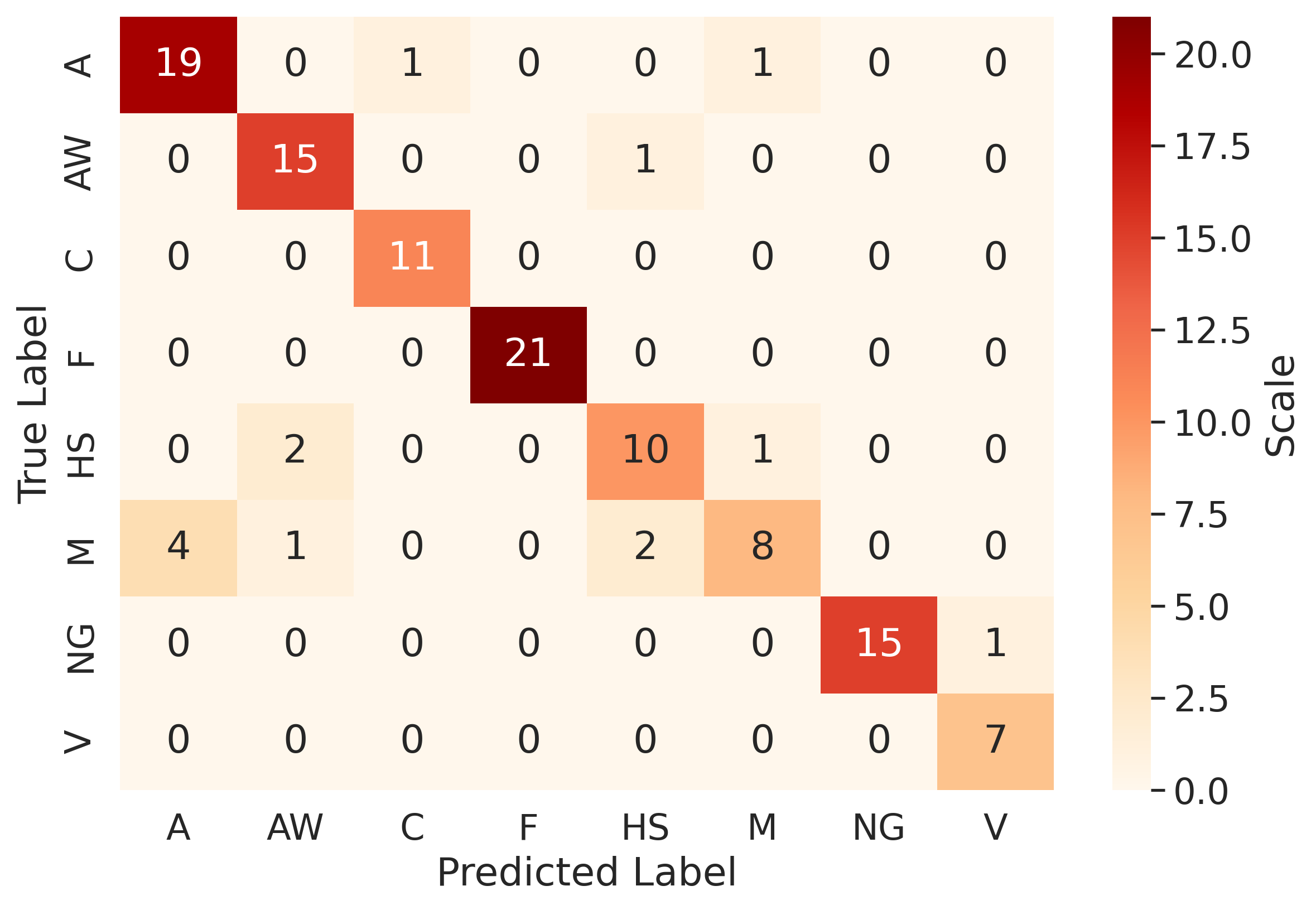}
     \caption{Confusion matrix of Random Forest classifier.}
    
    \end{subfigure}
    \caption{Confusion matrices of the two classifiers.}\label{FigConfusionMat}
\end{figure*}

We select the two models with the best accuracy performance, i.e. SVM and random forest, and provide further details to show how well our model performs in this specific task. Confusion matrices are provided in Figure~\ref{FigConfusionMat}. It shows how accurate the selected models are able to classify each of the victim accelerators. Values in each cell of the confusion matrix represent the number of samples of each (Predicted Label, True Label) pair. We can see from the figure that both models have acceptable accuracy performance ($69.2\%$ and $88.3\%$) and are able to differentiate the $8$ accelerator classes, although random forest works better with fewer misclassified samples and outperforms with a great margin. This could be due to the intrinsic features of the data traces, which are potentially more suitable for the algorithm of random forest and decision trees~\cite{goodfellow2016deep}.

\begin{figure}
    \centering
    \includegraphics[width=.95\linewidth]{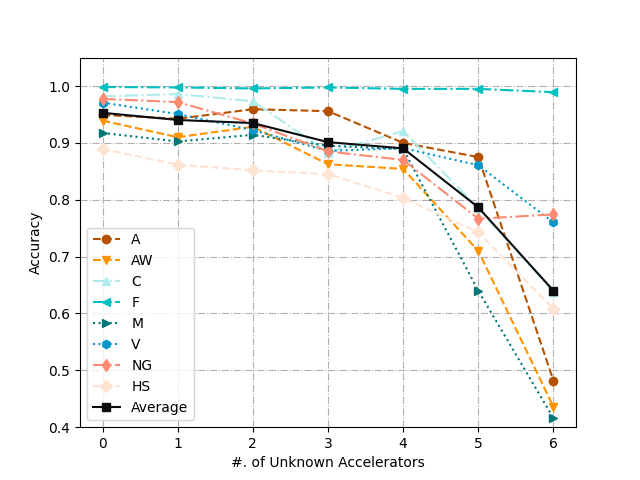}
    \caption{Open-world accuracy results.}
    \label{FigOpenWorldAccu}
\end{figure}

\vspace{10pt}
\noindent\textbf{Open-world.}
Then we also consider open-world fingerprinting scenario, where the attacker only has one specific target to recognize, and there are traces belonging to unseen accelerators during training process. During the experiments, we randomly select labels to remove and repeat the experiments multiple times to obtain the average accuracy performance data regarding classifiers corresponding to each class of victim accelerators. The accuracy results are shown in Figure~\ref{FigOpenWorldAccu}. We increment the number of unseen accelerators during training process and collect accuracy results for classifiers targeting different accelerators. We can see that the accuracy drops as the number of unknown accelerators increases. However, as long as the attacker has partial knowledge about accelerators in the system, when half of the traces are from unseen accelerators the fingerprinting accuracy can still maintain at around $80\%$ or higher. From Figure~\ref{FigOpenWorldAccu}, we can also see that ($1$) some accelerators are more vulnerable than others (e.g., our model on \texttt{fir} consistently achieves high fingerprinting success rate), highlighting the importance of providing protection when victim is \texttt{fir}; ($2$) when there are less types of accelerators, the fingerprinting success rate is higher and it is more important to provide defence.
%\textcolor{blue}{To justify the decrease when increase accelerators, you can mention that 1) some accelerators are more vulnerable to be identified by attackers; 2) when there are less types of accelerators running in cloud, it is more important to provide defenses.}

In the experiments above, we only use standard min-max scaling pre-processing and standard models. With further customization (filtering data, modifying predictive models), the accuracy can be potentially higher, resulting in a higher success rate and lower costs for continuing side-channel attacks.

\textbf{Summary.}
Our benchmark accelerator is able to capture communication patterns of co-located accelerators on FPGA and use these generated fingerprints to classify at a higher accuracy. This accuracy performance is sufficient for use in a real-world scenario. From the classifier side, we find that Random forest model achieves the highest classification accuracy and can reach $88\%$ classification accuracy. Surprisingly, the most complicated model, 1D-Convolution, achieves the worst classification accuracy performance. In our experiments, simpler models like random forest and SVM achieve significantly better accuracy performance. This could be potentially attributed to the limited number of traces in our data set.

\subsubsection{RQ $2$:}
As mentioned in Section~\ref{SecMethod}, our benchmark has $4$ different design parameter:
\begin{itemize}
    \item \texttt{ACCESS\_NUM}, which corresponds to how much traffic is generated by benchmark accelerator.
    \item \texttt{REPEAT\_NUM}, which is the number of times the kernel is executed and it relates to our measuring granularity and data stability.
    \item \texttt{BUFFER\_SIZE}, which determines how large each buffer is.
    \item \texttt{BUFFER\_NUM}, which relates to how many data points are collected within one performance trace.
\end{itemize}
The following parameter settings:
\begin{itemize}
    \item \texttt{ACCESS\_NUM}$ =1000$,
    \item \texttt{REPEAT\_NUM}$ =10$,
    \item \texttt{BUFFER\_SIZE}$ =4$ Bytes,
    \item \texttt{BUFFER\_NUM}$ =100$.
\end{itemize}
will be later referred to as our \textit{default setting}.

\begin{figure*}[ht!]
\centering
% \hfill
     \begin{subfigure}[t]{0.24\linewidth}
         \centering
         \includegraphics[width=\linewidth]{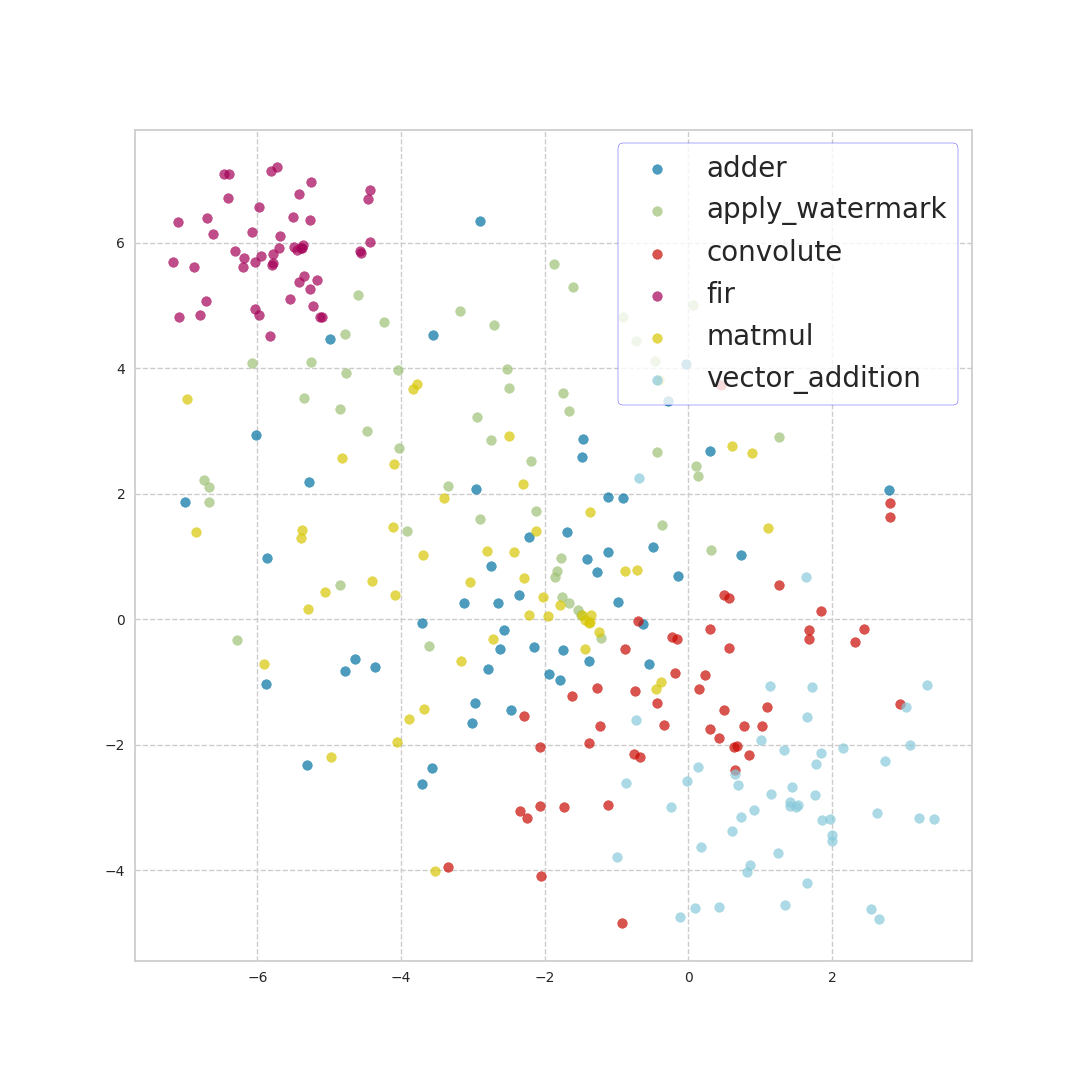}
         \caption{\texttt{ACCESS\_NUM} $ =250$.}
        %  \label{FigInstNum:b}
     \end{subfigure}
% \hfill
     \begin{subfigure}[t]{0.24\linewidth}
         \centering
         \includegraphics[width=\linewidth]{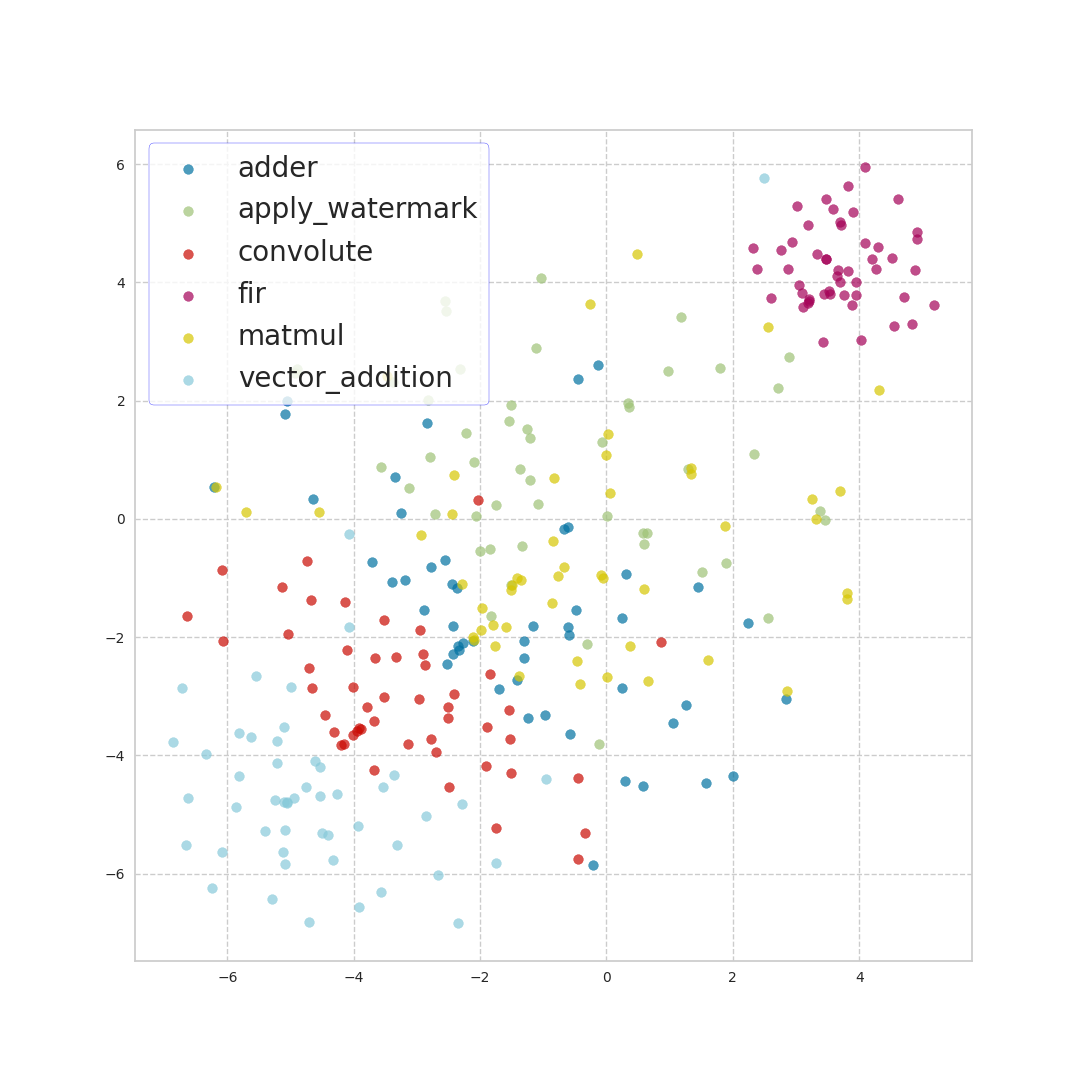}
         \caption{\texttt{ACCESS\_NUM} $ =500$.}
        %  \label{FigInstNum:b}
     \end{subfigure}
%  \hfill
%  \begin{subfigure}[t]{0.24\linewidth}
%      \centering
%      \includegraphics[width=\linewidth]{Figs/access1000.png}
%      \caption{\texttt{ACCESS\_NUM} $ =1000$.}
%     %  \label{FigInstNum:b}
%  \end{subfigure}
% \hfill
 \begin{subfigure}[t]{0.24\linewidth}
     \centering
     \includegraphics[width=\linewidth]{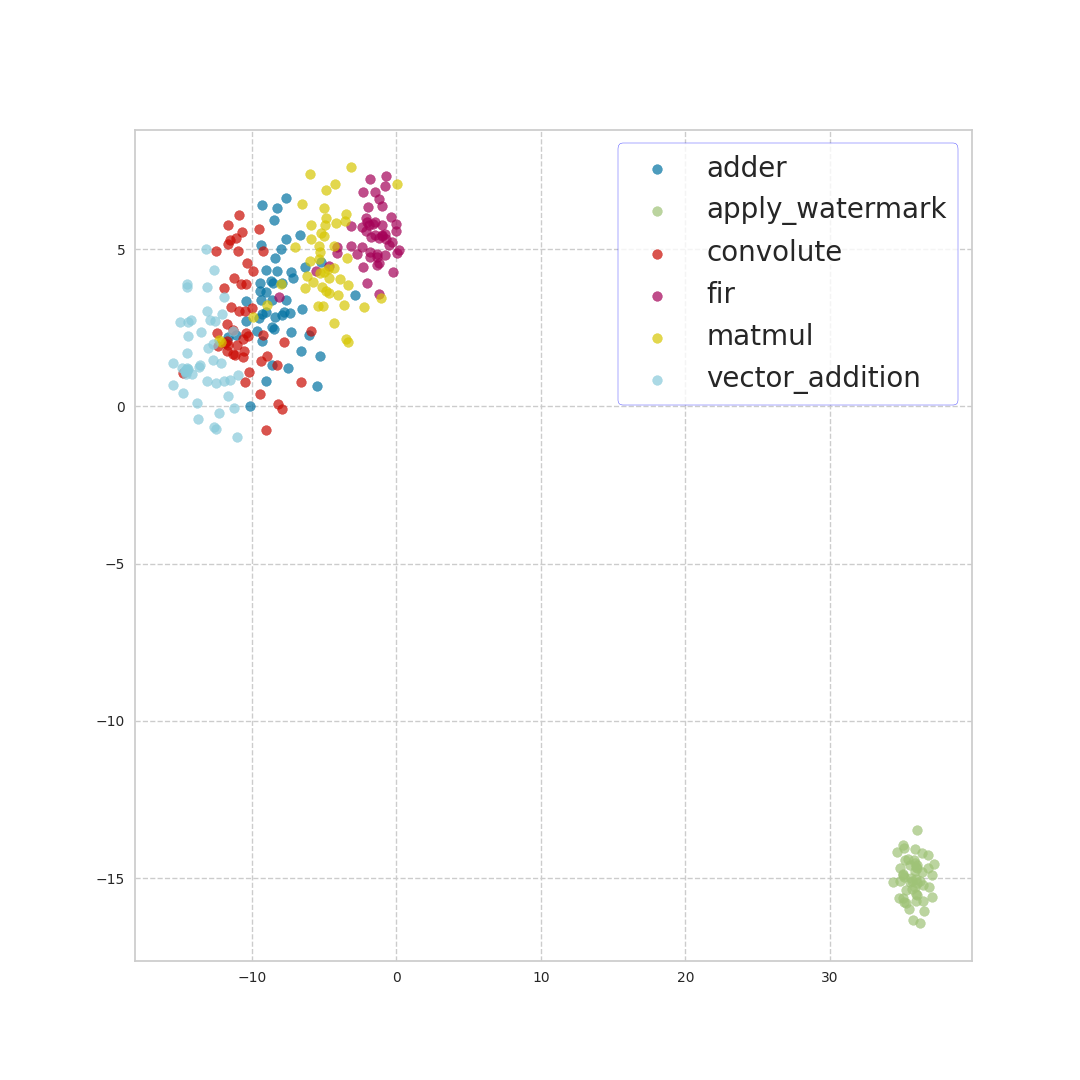}
     \caption{\texttt{ACCESS\_NUM} $ =2000$.}
    %  \label{FigInstNum:b}
 \end{subfigure}
% \hfill
 \begin{subfigure}[t]{0.24\linewidth}
     \centering
     \includegraphics[width=\linewidth]{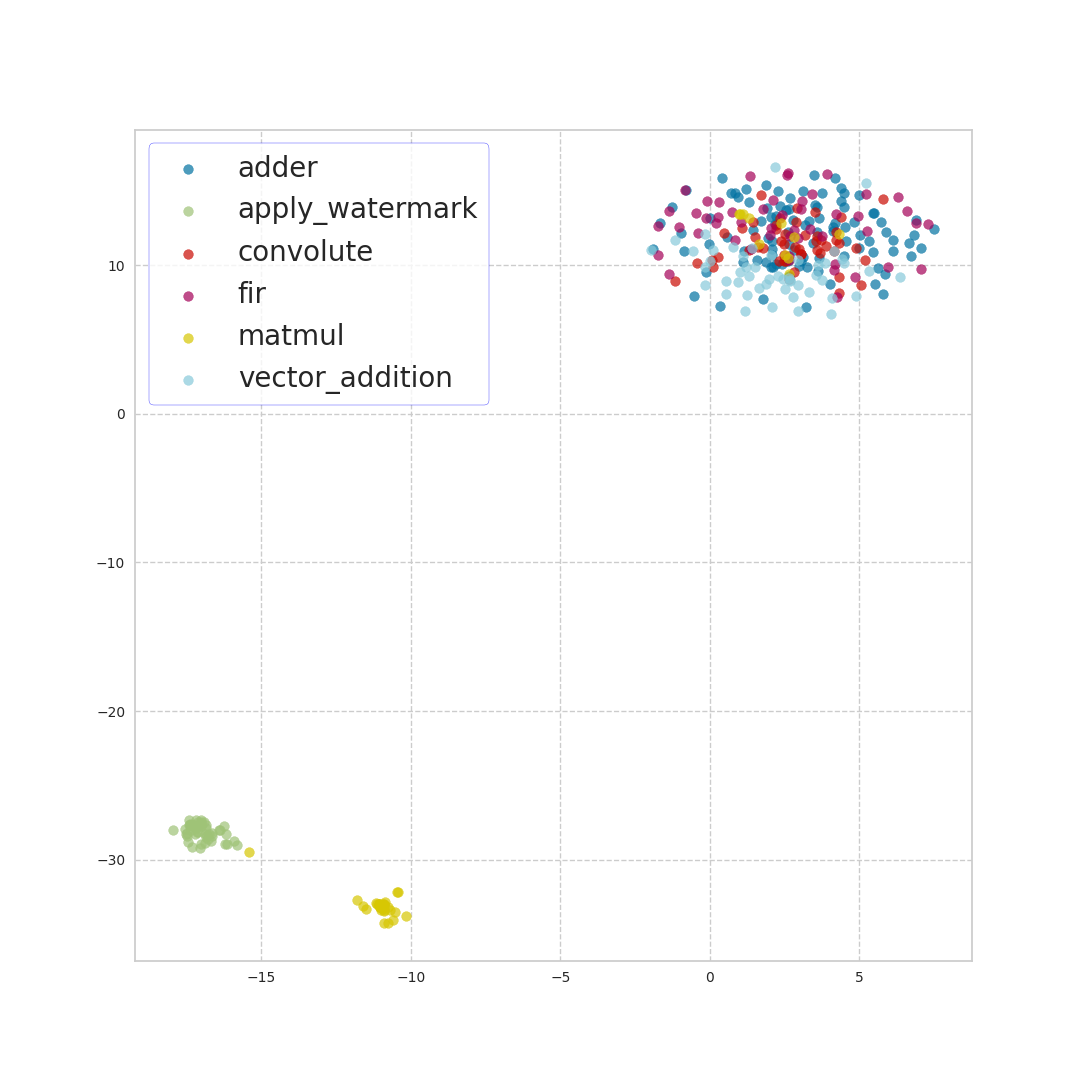}
     \caption{\texttt{ACCESS\_NUM} $ =4000$.}
    %  \label{FigInstNum:b}
 \end{subfigure}
% \hfill

\caption{Traces and corresponding accuracy results under different \texttt{ACCESS\_NUM} settings.}\label{FigACCESSNUM}
\end{figure*}

\begin{figure*}[ht!]
    \centering
    % \hfill
         \begin{subfigure}[t]{0.24\linewidth}
             \centering
             \includegraphics[width=\linewidth]{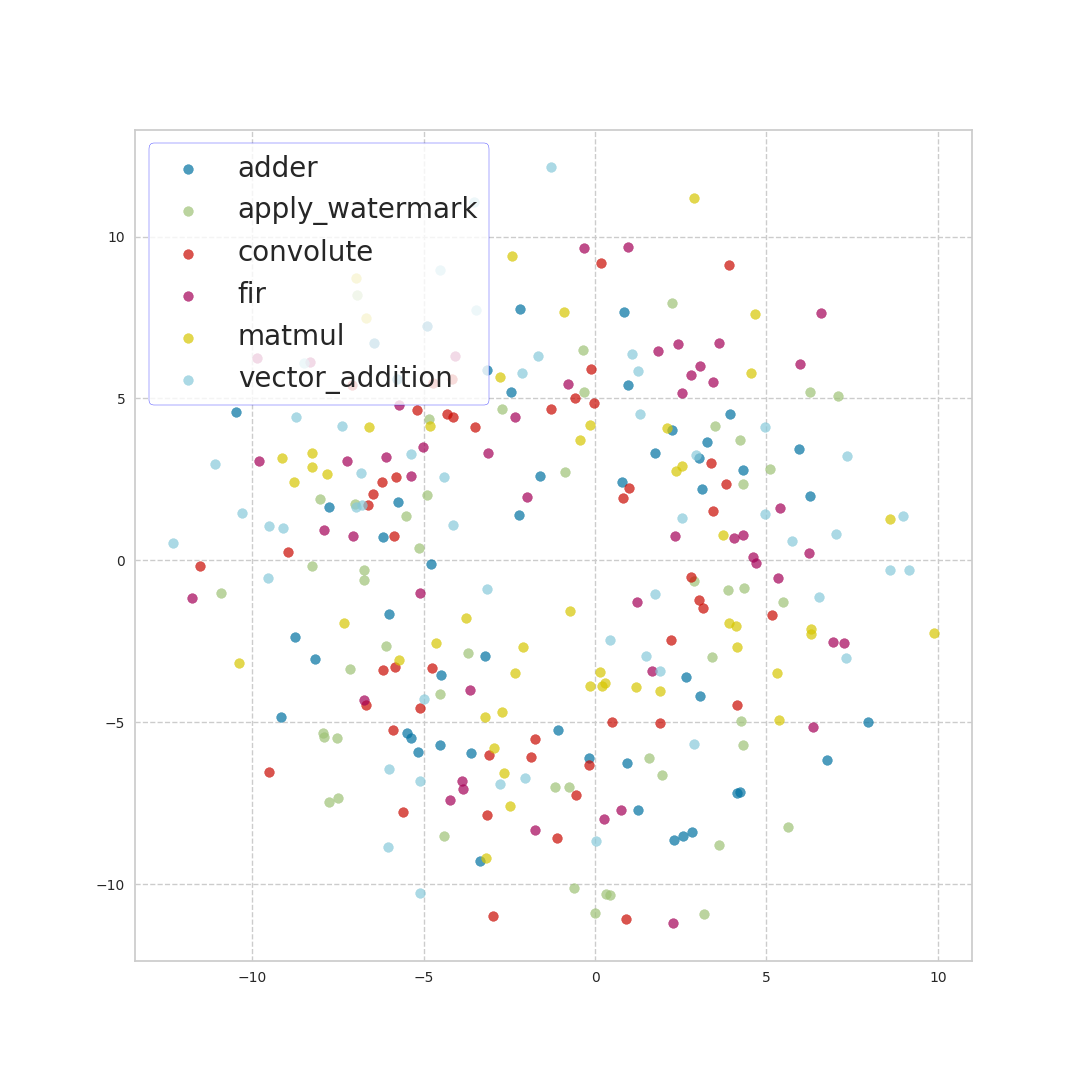}
             \caption{\texttt{REPEAT\_NUM} = $1$.}
            %  \label{FigInstNum:b}
         \end{subfigure}
    % \hfill
         \begin{subfigure}[t]{0.24\linewidth}
             \centering
             \includegraphics[width=\linewidth]{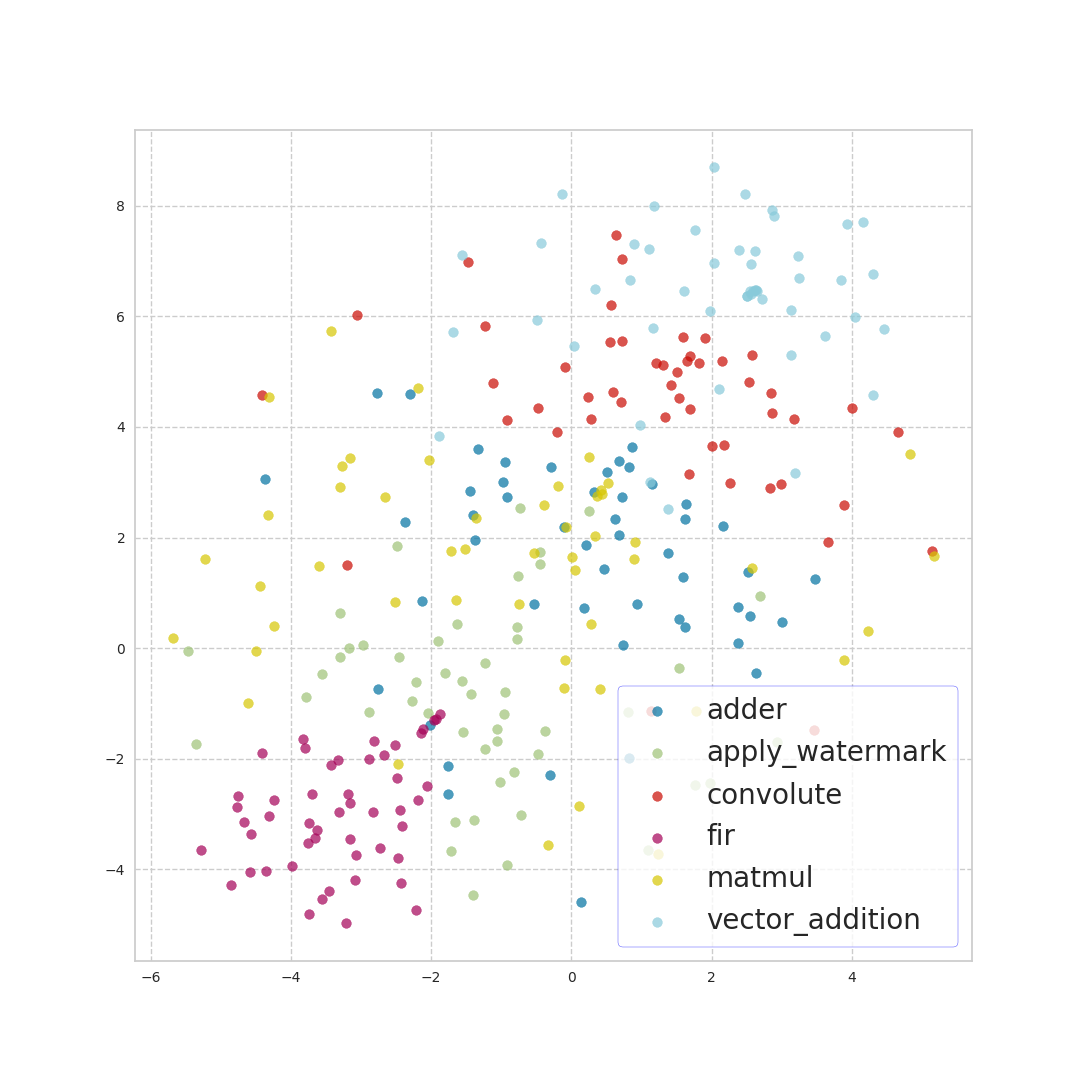}
             \caption{\texttt{REPEAT\_NUM} = $5$.}
            %  \label{FigInstNum:b}
         \end{subfigure}
    %  \hfill
    %  \begin{subfigure}[t]{0.24\linewidth}
    %      \centering
    %      \includegraphics[width=\linewidth]{Figs/repeat10.png}
    %      \caption{\texttt{REPEAT\_NUM} = $10$ Bytes.}
    %     %  \label{FigInstNum:b}
    %  \end{subfigure}
    % \hfill
     \begin{subfigure}[t]{0.24\linewidth}
         \centering
         \includegraphics[width=\linewidth]{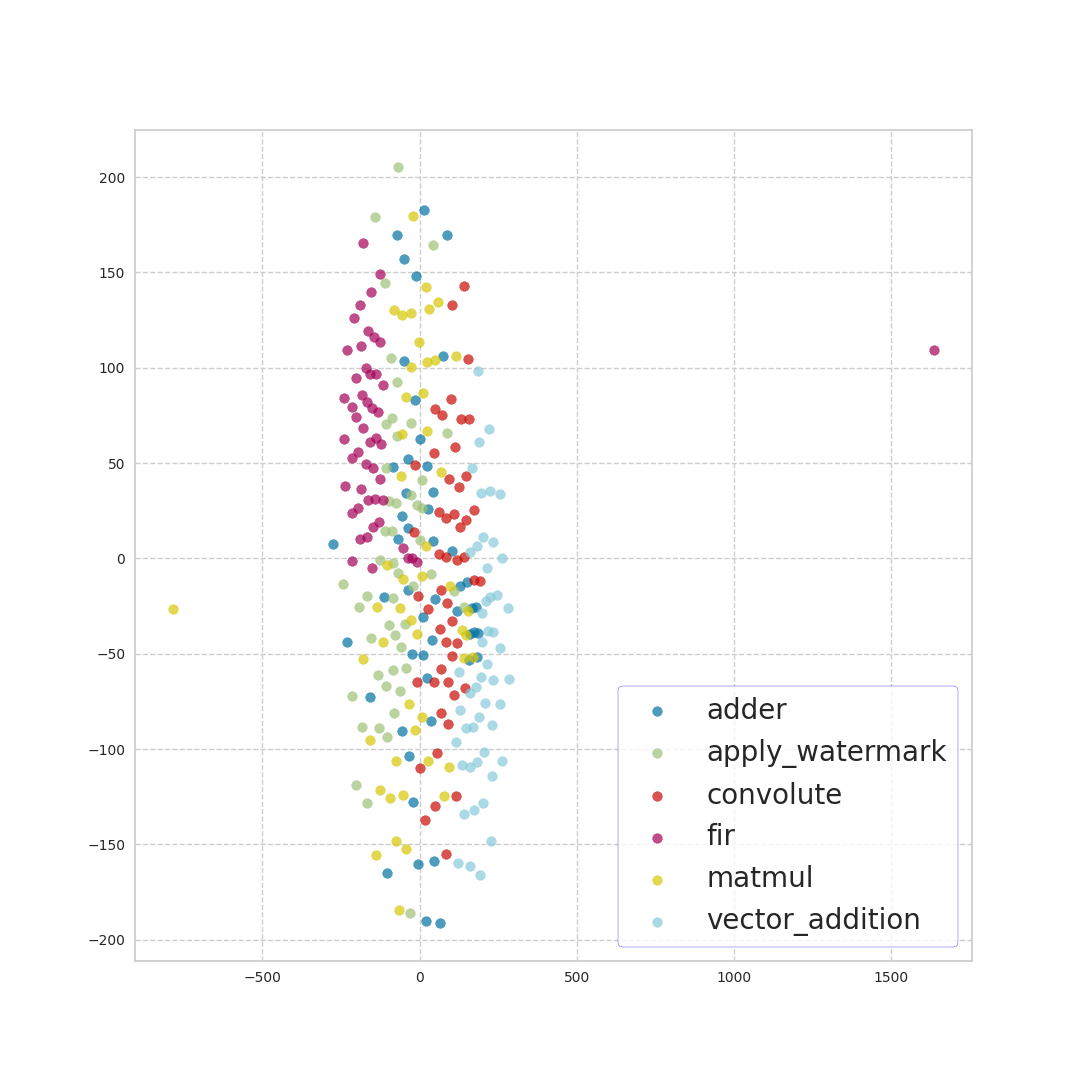}
         \caption{\texttt{REPEAT\_NUM} = $20$.}
        %  \label{FigInstNum:b}
     \end{subfigure}
    % \hfill
     \begin{subfigure}[t]{0.24\linewidth}
         \centering
         \includegraphics[width=\linewidth]{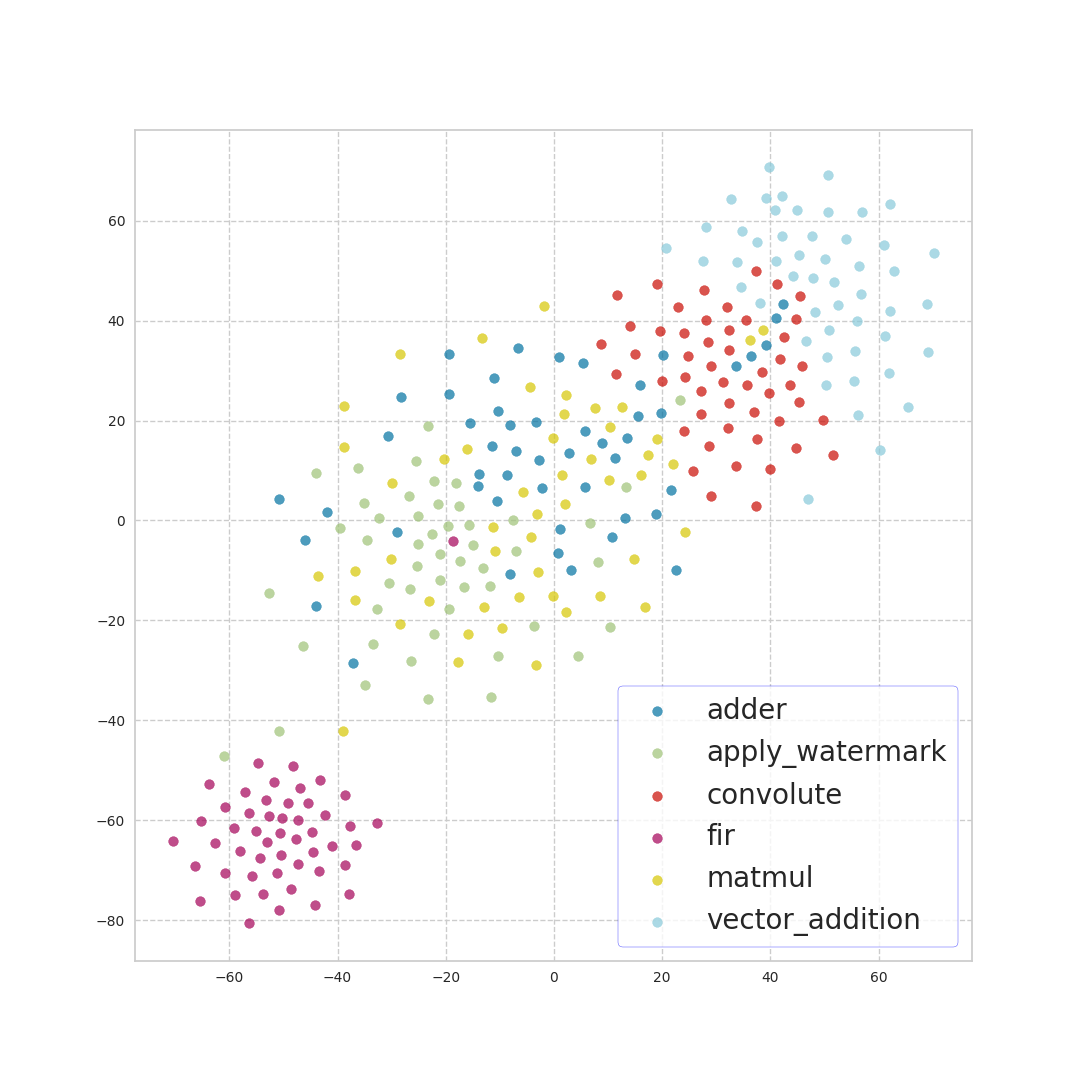}
         \caption{\texttt{REPEAT\_NUM} = $50$.}
        %  \label{FigInstNum:b}
     \end{subfigure}
    % \hfill

    \caption{Traces and corresponding accuracy results under different \texttt{REPEAT\_NUM} settings.}\label{FigREPEATNUM}
    \end{figure*}

\begin{figure*}[ht!]
\centering
% \hfill
     \begin{subfigure}[t]{0.24\linewidth}
         \centering
         \includegraphics[width=\linewidth]{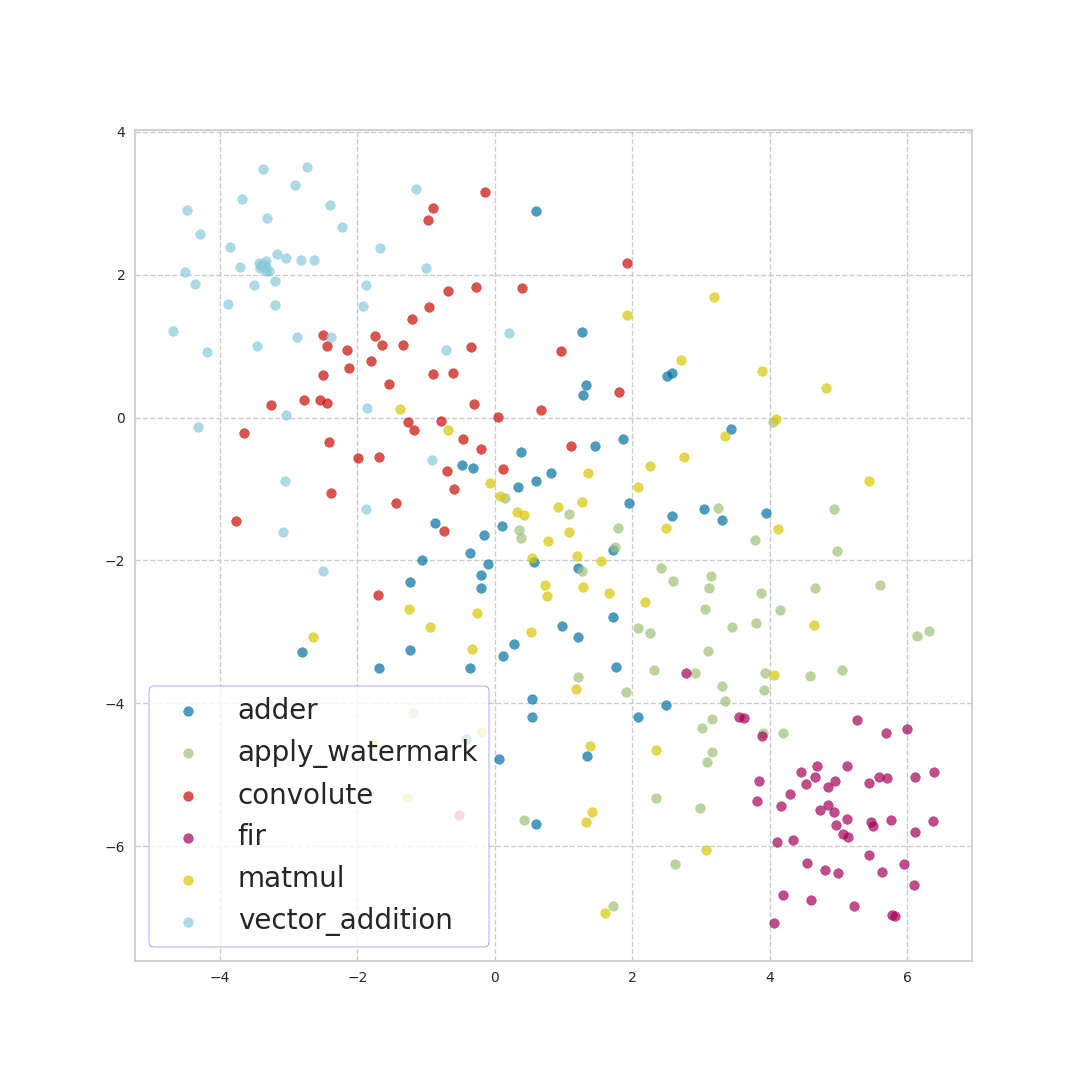}
         \caption{\texttt{BUFFER\_SIZE} = $1$ Bytes.}
        %  \label{FigInstNum:b}
     \end{subfigure}
% \hfill
     \begin{subfigure}[t]{0.24\linewidth}
         \centering
         \includegraphics[width=\linewidth]{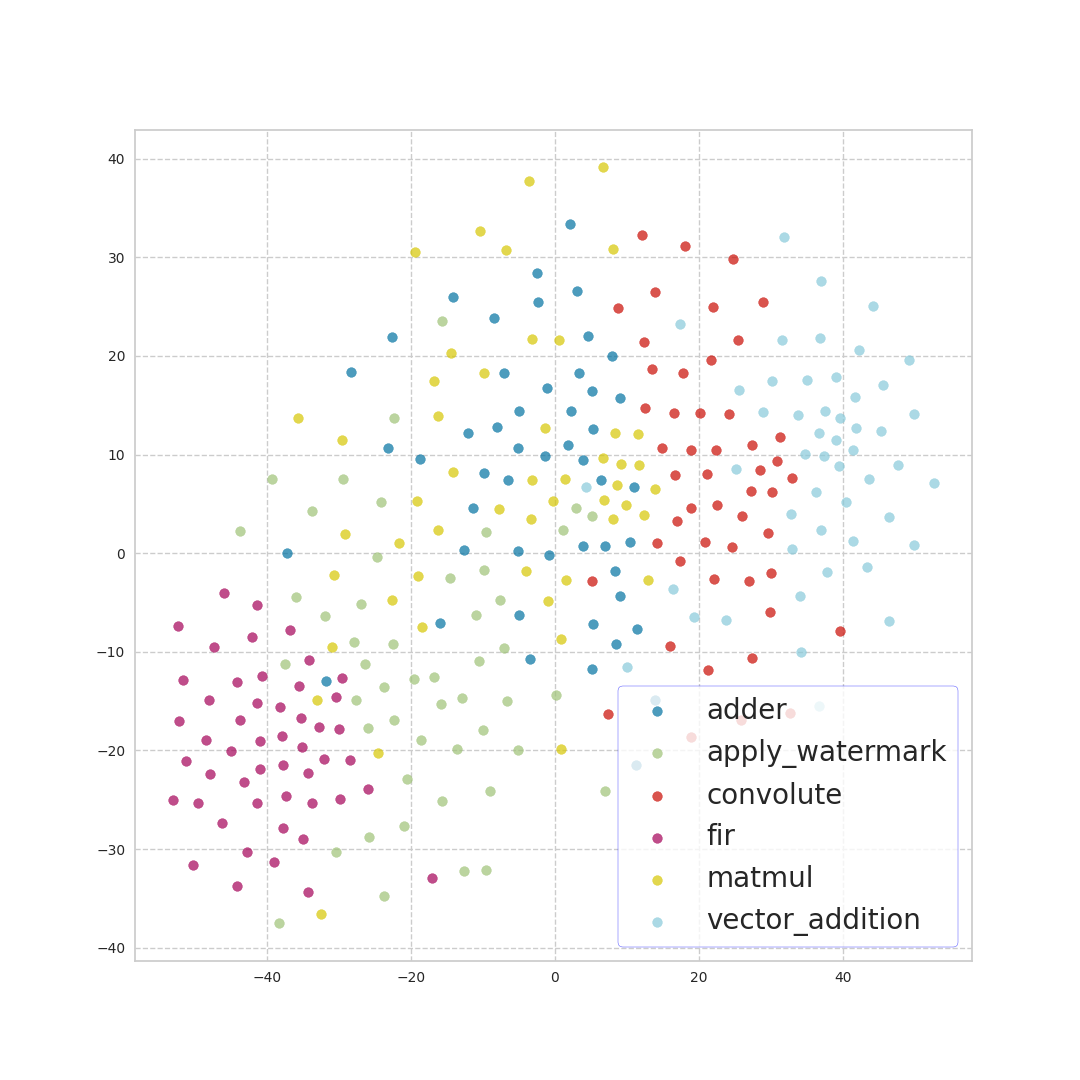}
         \caption{\texttt{BUFFER\_SIZE} = $2$ Bytes.}
        %  \label{FigInstNum:b}
     \end{subfigure}
%  \hfill
%  \begin{subfigure}[t]{0.24\linewidth}
%      \centering
%      \includegraphics[width=\linewidth]{Figs/chunk4.png}
%      \caption{\texttt{BUFFER\_SIZE} = $4$ Bytes.}
%     %  \label{FigInstNum:b}
%  \end{subfigure}
% \hfill
 \begin{subfigure}[t]{0.24\linewidth}
     \centering
     \includegraphics[width=\linewidth]{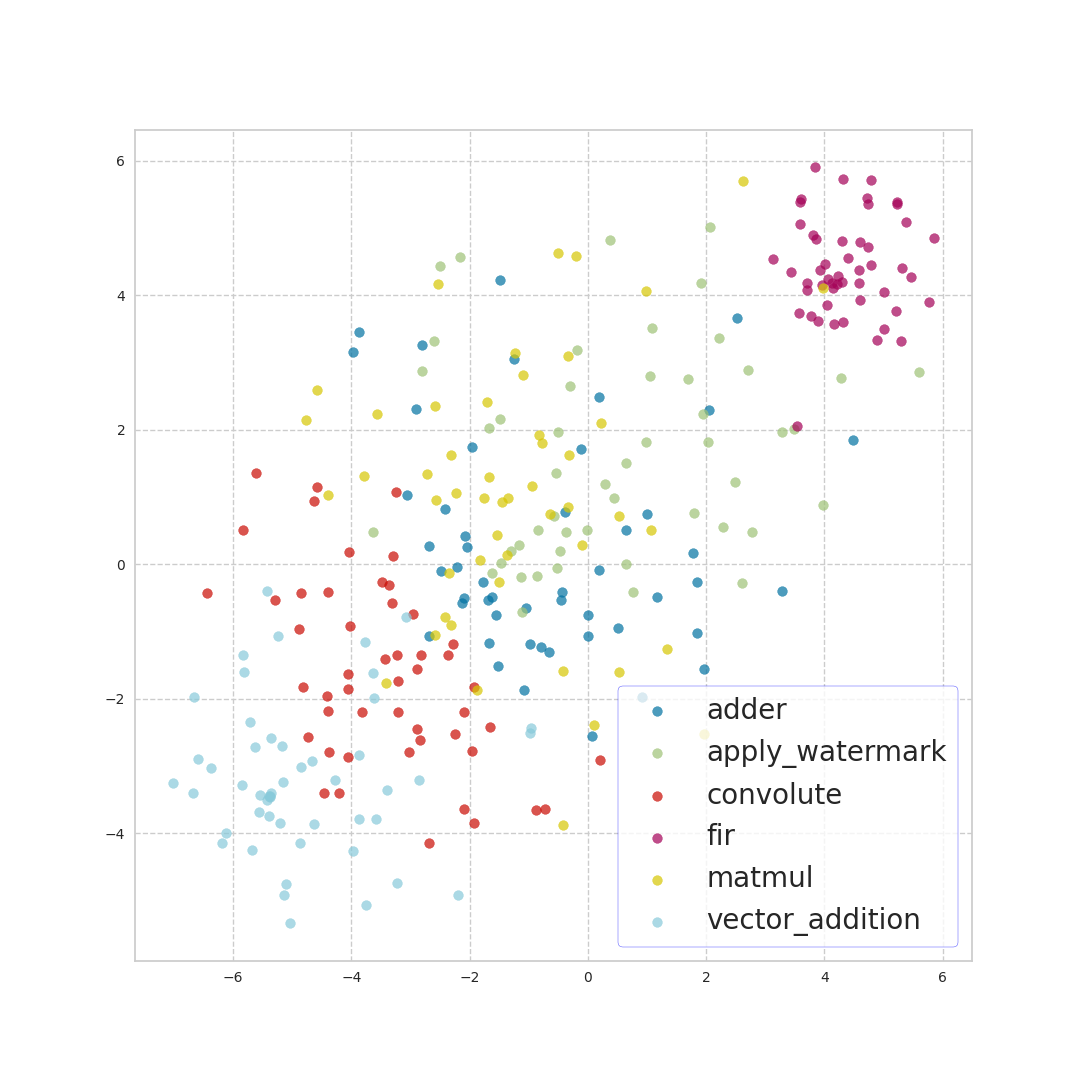}
     \caption{\texttt{BUFFER\_SIZE} = $8$ Bytes.}
    %  \label{FigInstNum:b}
 \end{subfigure}
% \hfill
 \begin{subfigure}[t]{0.24\linewidth}
     \centering
     \includegraphics[width=\linewidth]{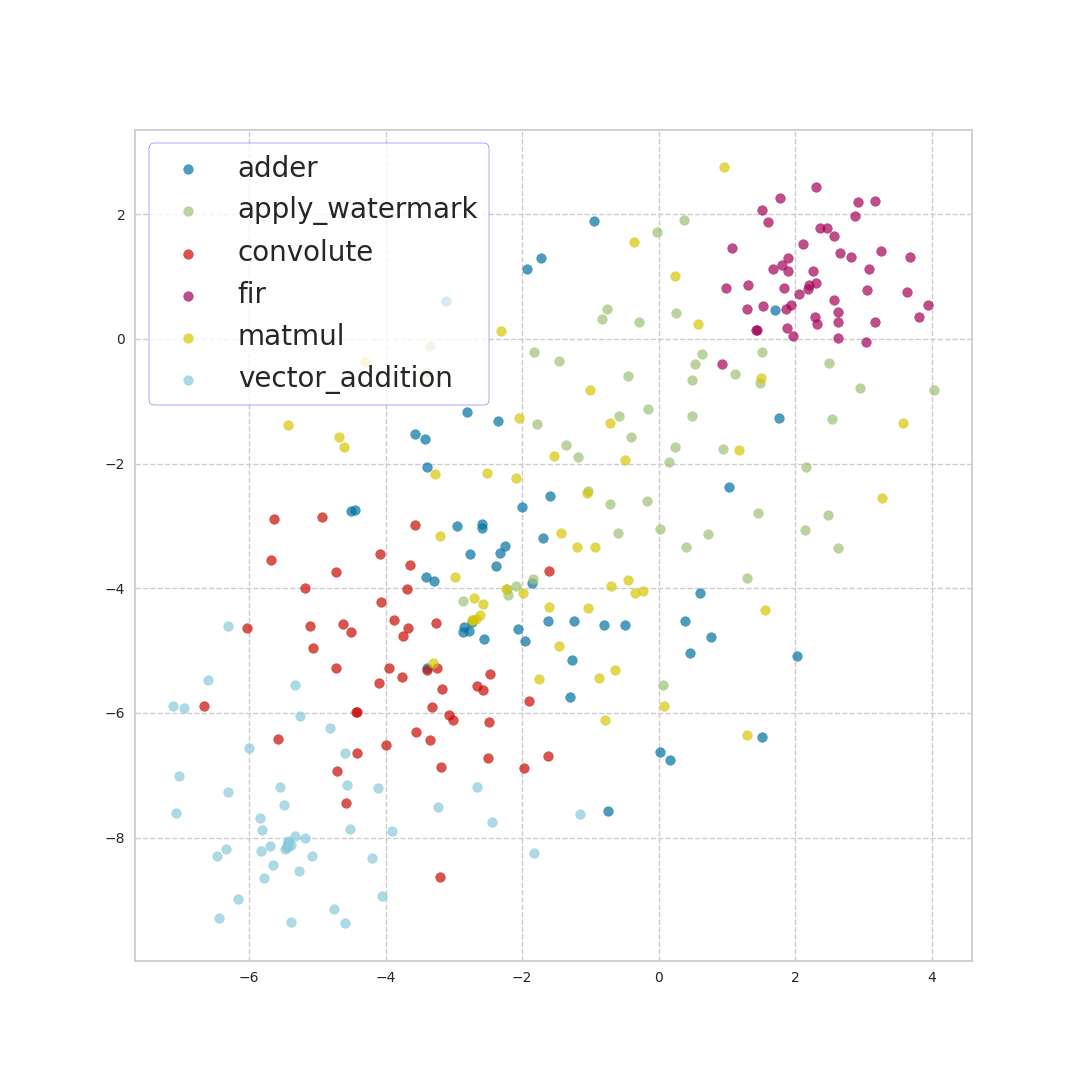}
     \caption{\texttt{BUFFER\_SIZE} = $16$ Bytes.}
    %  \label{FigInstNum:b}
 \end{subfigure}
% \hfill

\caption{Traces and corresponding accuracy results under different \texttt{BUFFER\_SIZE} settings.}\label{FigBUFFERSIZE}
\end{figure*}

\begin{figure*}[ht!]
\centering
% \hfill
     \begin{subfigure}[t]{0.24\linewidth}
         \centering
         \includegraphics[width=\linewidth]{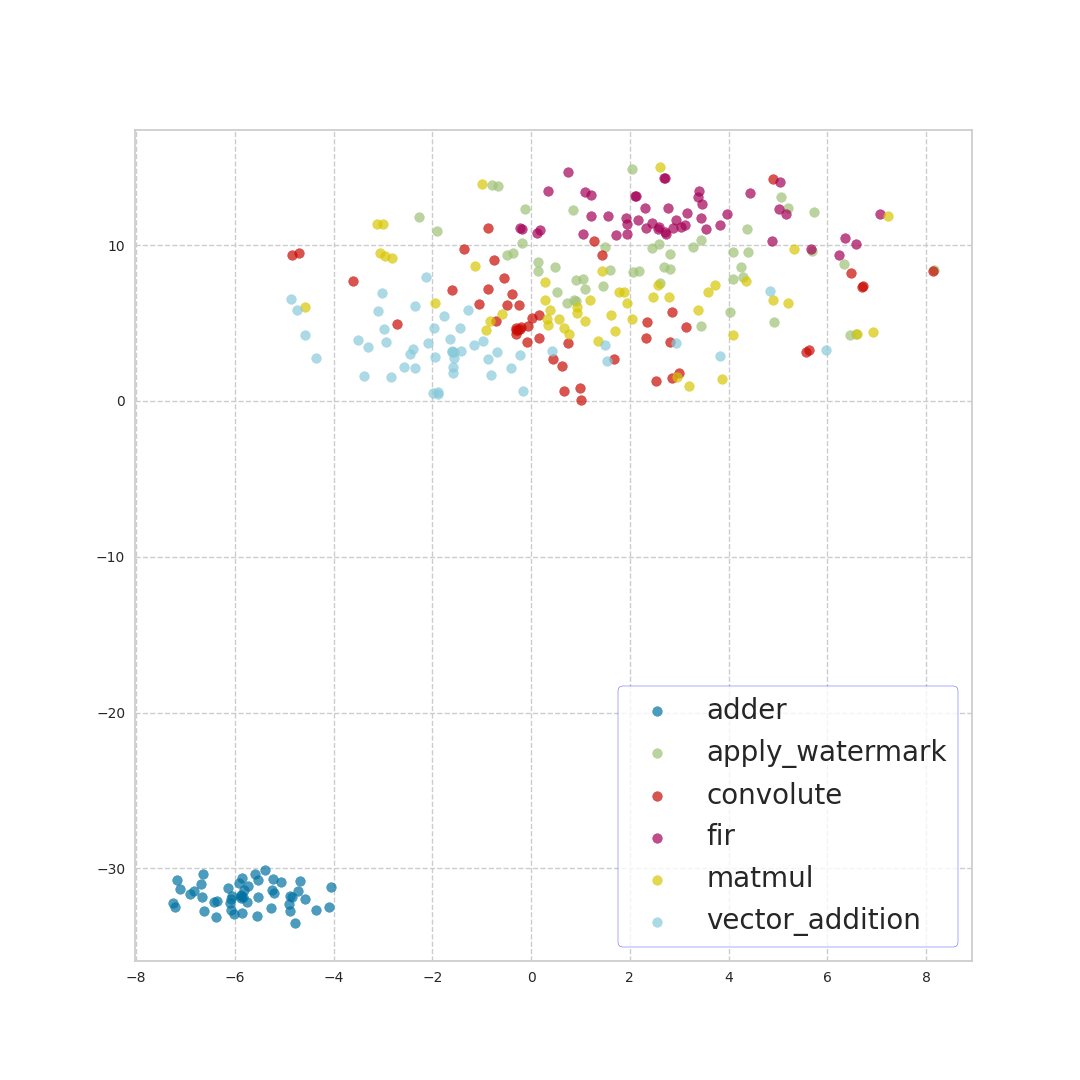}
         \caption{\texttt{BUFFER\_NUM} = $50$.}
        %  \label{FigInstNum:b}
     \end{subfigure}
% \hfill
    %  \begin{subfigure}[t]{0.24\linewidth}
    %      \centering
    %      \includegraphics[width=\linewidth]{Figs/buffer100.png}
    %      \caption{\texttt{BUFFER\_NUM} = $100$.}
    %     %  \label{FigInstNum:b}
    %  \end{subfigure}
%  \hfill
 \begin{subfigure}[t]{0.24\linewidth}
     \centering
     \includegraphics[width=\linewidth]{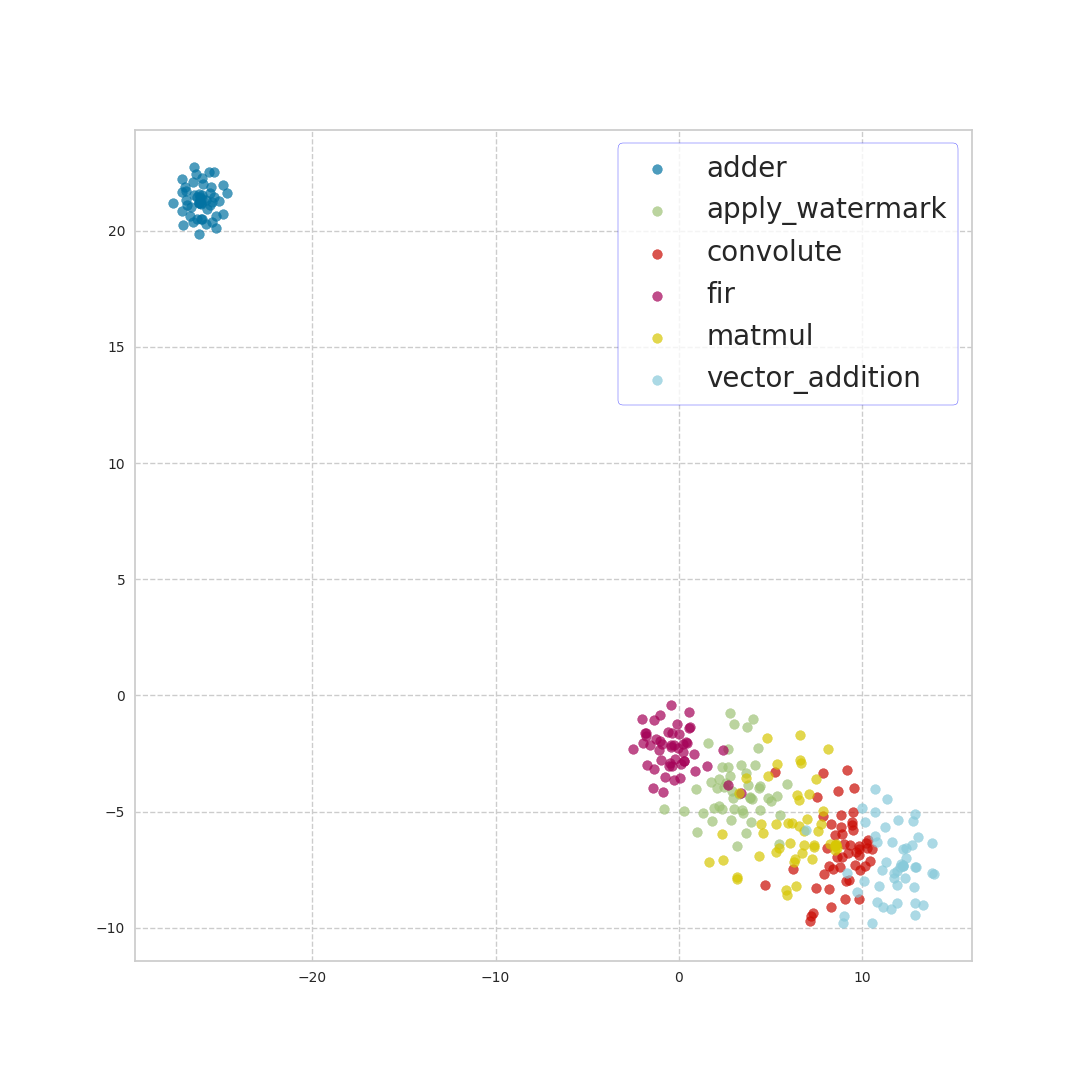}
     \caption{\texttt{BUFFER\_NUM} = $200$.}
    %  \label{FigInstNum:b}
 \end{subfigure}
% \hfill
 \begin{subfigure}[t]{0.24\linewidth}
     \centering
     \includegraphics[width=\linewidth]{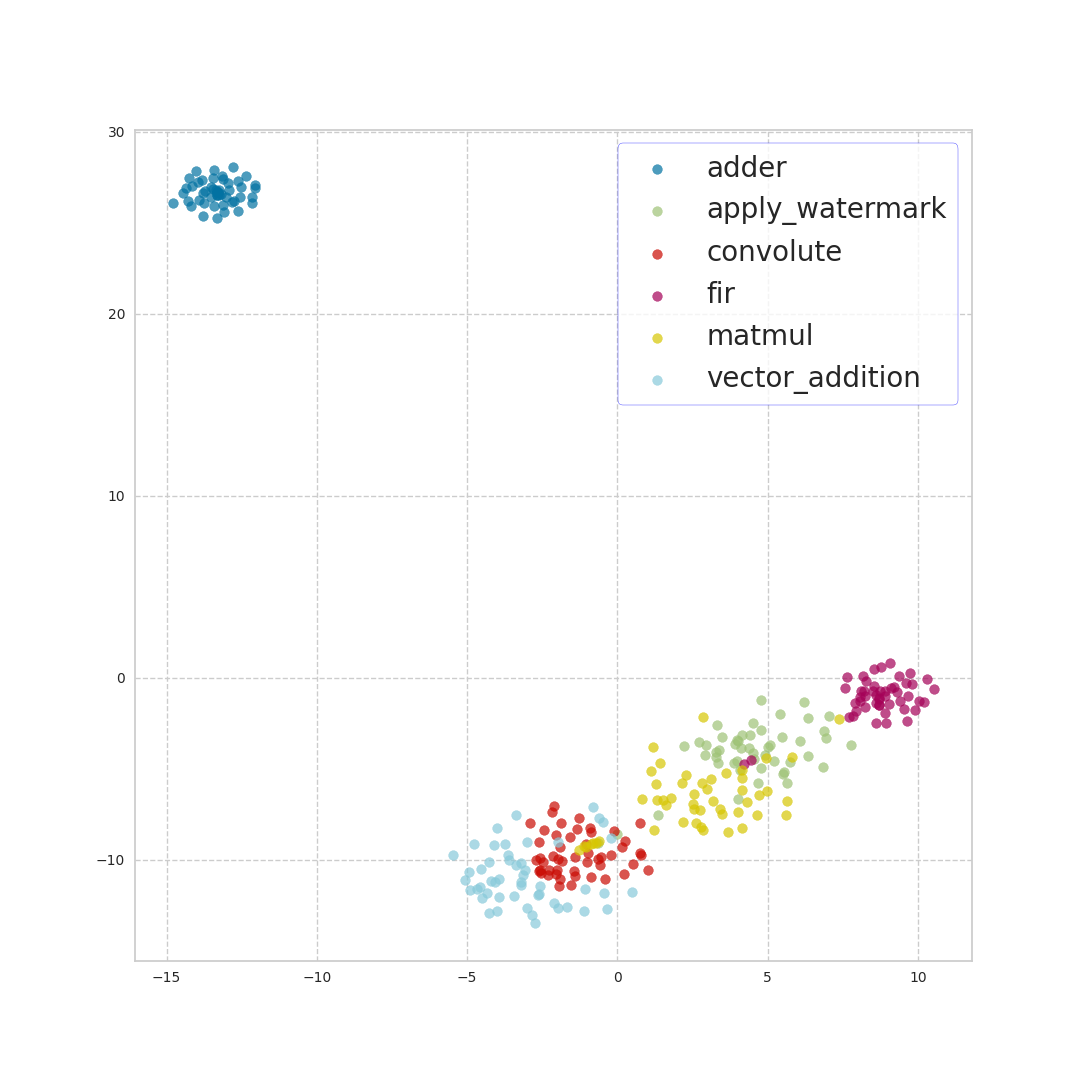}
     \caption{\texttt{BUFFER\_NUM} = $400$.}
    %  \label{FigInstNum:b}
 \end{subfigure}
% \hfill
 \begin{subfigure}[t]{0.24\linewidth}
     \centering
     \includegraphics[width=\linewidth]{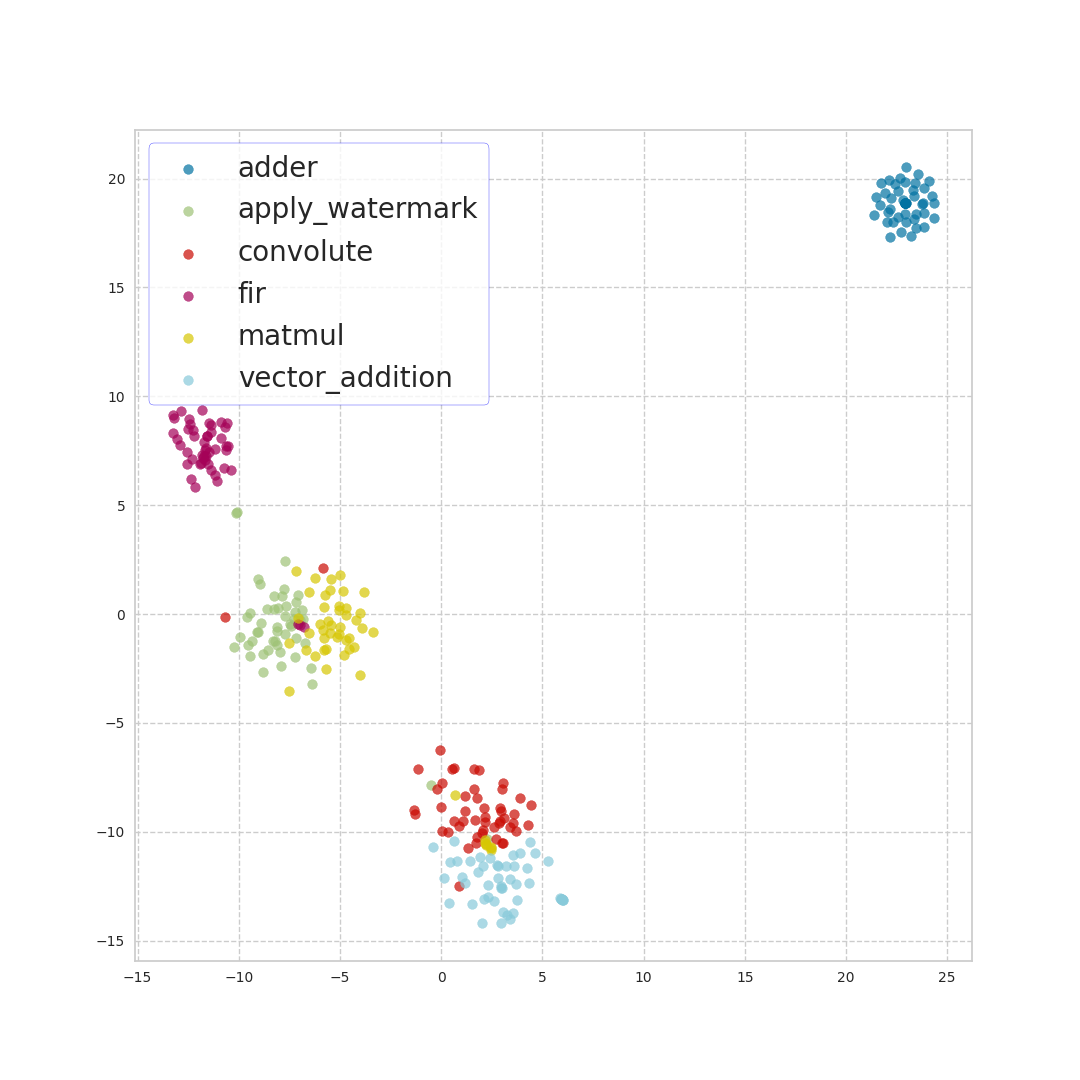}
     \caption{\texttt{BUFFER\_NUM} = $800$.}
    %  \label{FigInstNum:b}
 \end{subfigure}
% \hfill

\caption{Traces and corresponding accuracy results under different \texttt{BUFFER\_NUM} settings.}\label{FigBUFFERNUM}
\end{figure*}

\begin{figure*}[ht!]
\centering
% \hfill

% \hfill
    %  \begin{subfigure}[t]{0.24\linewidth}
    %      \centering
    %      \includegraphics[width=\linewidth]{Figs/buffer100.png}
    %      \caption{\texttt{BUFFER\_NUM} = $100$.}
    %     %  \label{FigInstNum:b}
    %  \end{subfigure}
%  \hfill
 \begin{subfigure}[t]{0.24\linewidth}
     \centering
     \includegraphics[width=\linewidth]{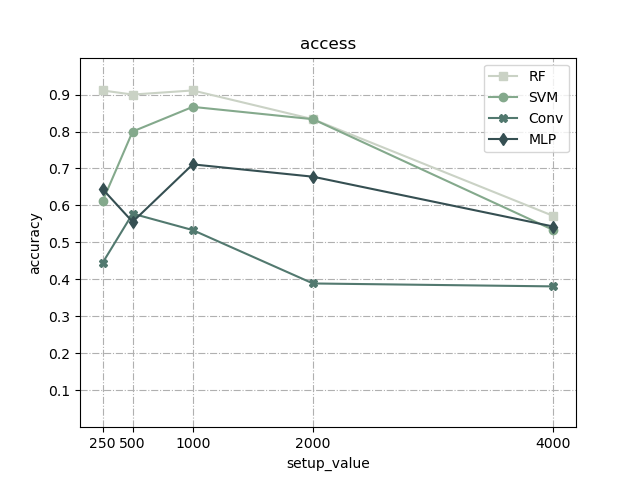}
     \caption{Varying \texttt{ACCESS\_NUM}.}
    %  \label{FigInstNum:b}
 \end{subfigure}
% \hfill
 \begin{subfigure}[t]{0.24\linewidth}
     \centering
     \includegraphics[width=\linewidth]{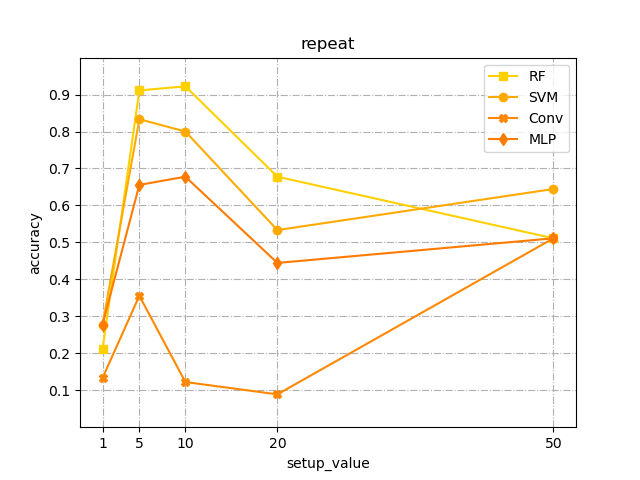}
     \caption{Varying \texttt{REPEAT\_NUM}.}
    %  \label{FigInstNum:b}
 \end{subfigure}
% \hfill
 \begin{subfigure}[t]{0.24\linewidth}
     \centering
     \includegraphics[width=\linewidth]{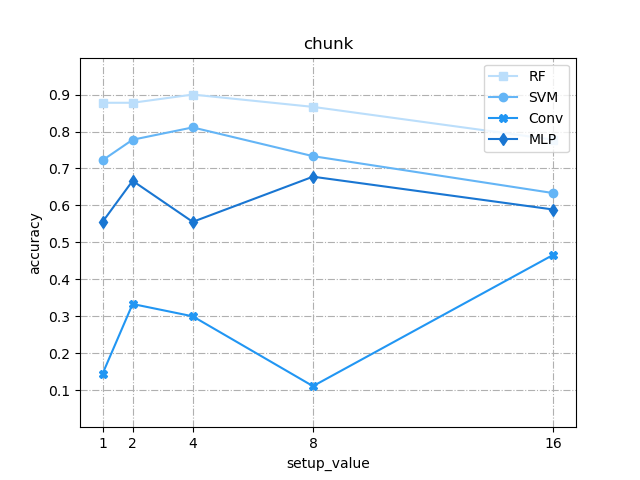}
     \caption{Varying \texttt{BUFFER\_SIZE}.}
    %  \label{FigInstNum:b}
 \end{subfigure}
% \hfill
 \begin{subfigure}[t]{0.24\linewidth}
     \centering
     \includegraphics[width=\linewidth]{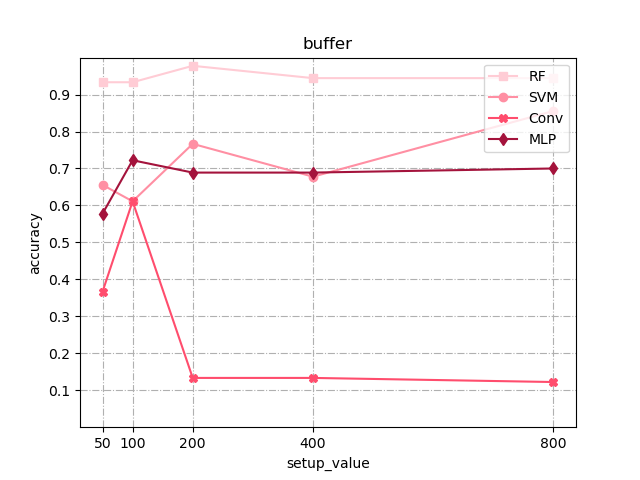}
     \caption{Varying \texttt{BUFFER\_NUM}.}
    %  \label{FigInstNum:b}
 \end{subfigure}
\caption{Traces and corresponding accuracy results under different \texttt{BUFFER\_NUM} settings.}\label{FigAccuracy}
\end{figure*}

In this experiment, we screen all parameters and provide t-SNE visualization and compare their classification accuracy with the one under the default setting. To make visualization results clearer, we drop the simplest accelerator (\texttt{noisegen}) and the most complex accelerator (\texttt{hotspot}) and only perform attacks on the remaining $6$ accelerators. %The screening process is based on , since this setting generates reasonable classification accuracy empirically. 
To explore the effects of each of the $4$ parameters, we fix the other $3$ parameters to the default settings and vary the value of the target parameter, then collect data on the $6$ victim accelerators. The t-SNE visualization of the collected data traces as well as classification accuracy traces of our $4$ models regarding the $4$ parameters are shown in Figure~\ref{FigACCESSNUM} - \ref{FigBUFFERNUM}. Corresponding accuracy performance of the $4$ models are provided in Figure~\ref{FigAccuracy}. Figures corresponding to configurations that are identical with default settings are omitted to avoid repetition, the results are the same as in Figure~\ref{FigTrace}~(b). In these figures, we obtain t-SNE results from normalized communication bandwidth data. Overall, the use of different parameter values results in varying trace patterns and can hence affect the accuracy of different models. The analysis of the results we obtain in this experiment is shown as follows.

\textbf{\texttt{ACCESS\_NUM}.}
In Figure~\ref{FigACCESSNUM}, the influence of the parameter \texttt{ACCESS \_NUM} is shown. From Figure~\ref{FigACCESSNUM}~(a) - (d), we can observe a change in the visualization results, i.e. the traces collected from different victim accelerators show different separability. This indicates that the ability of our accelerator benchmark to capture the differentiable patterns in I/O operations existing in our victim accelerators can vary according to \texttt{ACCESS\_NUM}. We can see that after \texttt{ACCESS\_NUM} $\geq 2000$, data points from several accelerator classes tend to be mixed together. By looking at Figure~\ref{FigAccuracy}~(a), we can see that the random forest model achieves the highest accuracy result, reaching an accuracy over $90\%$ at \texttt{ACCESS\_NUM}$ =1000$. SVM achieves over $85\%$ accuracy and MLP achieves over $70\%$ classification accuracy, both at the same point. However, the 1D-Convolution model is only able to achieve $60\%$ accuracy at its highest. We can also see that as \texttt{ACCESS\_NUM} increases, except 1D-Convolution model, the accuracy generally increases first (though the accuracy of our random forest model stays over $90\%$ relatively stably). After reaching the highest accuracy at \texttt{ACCESS\_NUM}$ =1000$, the accuracy starts to drop.

We speculate that the reason behind the fingerprinting accuracy difference is due to the measurement granularity differences when \texttt{ACCESS\_NUM} ranges from $250$ to $4000$. Initially, the growth of \texttt{ACCESS\_NUM} introduces more data to be read and written hence the effects of noise can be better cancelled and communication patterns can be better captured until \texttt{ACCESS\_NUM} reaches $1000$. However, since the execution time of our benchmark accelerator also increases as \texttt{ACCESS\_NUM} grows, the measurement will become more coarse-grained since the change in I/O performance variance of victim accelerators within this execution time period will be amortized. After a certain point (in our experiment, between $1000$ and $2000$), the extended execution time of benchmark accelerator causes the benchmark circuit to lose the ability to accurately capture victim communication patterns, thereby inducing a drop in classification accuracy.

\textbf{\texttt{REPEAT\_NUM.}}
For parameter \texttt{REPEAT\_NUM}, the t-SNE visualization results are shown in Figure~\ref{FigREPEATNUM}. Same as in Figure~\ref{FigACCESSNUM}, data points belonging to different accelerator classes in Figure~\ref{FigREPEATNUM}~(a) - (d) are separable, where the clearest clustering results appear at \texttt{REPEAT\_NUM}$ =5$ and \texttt{REPEAT\_NUM}$ =10$ (see Figure~\ref{FigTrace}). The classification results for the machine learning models in Figure~\ref{FigAccuracy}~(b) also match this observation, with the highest classification results achieved at \texttt{REPEAT\_NUM}$ =5$ and \texttt{REPEAT\_NUM}$ =10$, where the accuracy of random forest is again over $90\%$ and the highest accuracy results of SVM and MLP are around $85\%$ and $70\%$, respectively. In Figure~\ref{FigAccuracy}~(b) we can observe a similar trend as in Figure~\ref{FigAccuracy}~(a), where the accuracy first increases to an optimal point and starts to drop as \texttt{REPEAT\_NUM} grows.

The explanation for the accuracy trend is similar. As \texttt{REPEAT\_NUM} determines how many times our benchmark is executed when operating on a memory buffer, increasing \texttt{REPEAT\_NUM} will: ($1$) cancel the effects of noise and obtain a more precise measurement of the performance; ($2$) extend the time it takes to operate on a single buffer, i.e. the time it takes to generate a data point in the performance trace. As the execution time of the accelerator kernel task is relatively short, when \texttt{REPEAT\_NUM} is low, the measurement will be finished within a short period of time and the dynamic communication patterns cannot be captured by our benchmark accelerator. This, combining the influence of noise, is the reason why all models achieve poor classification accuracy results at \texttt{REPEAT\_NUM}$ =1$. When \texttt{REPEAT\_NUM} increases, the communication patterns start to be captured. However, if \texttt{REPEAT\_NUM} is too large, same as the situation in \texttt{ACCESS\_NUM} the whole measurement process becomes too coarse-grained. Changes in the I/O communication traffic may be amortized, thus classification models cannot extract detailed information from the collected traces.

\textbf{\texttt{BUFFER\_SIZE}.} The experimental results of trace visualization and classification results under different \texttt{BUFFER\_SIZE} settings are shown in Figure~\ref{FigBUFFERSIZE}. We can see that all the \texttt{BUFFER\_SIZE} settings we use are able to preserve the layering information in the victim accelerators. From Figure~\ref{FigAccuracy}~(c), we also observe that the influence of parameter \texttt{BUFFER\_SIZE} is not as much as \texttt{ACCESS\_NUM} and \texttt{REPEAT\_NUM}. However, there is an optimal point for the random forest and SVM to work on (\texttt{BUFFER\_SIZE}$ =4$ Bytes). We will keep using this empirical value since it is the best work point for our most accurate model. However, as \texttt{BUFFER\_SIZE} increases beyond the optimal point, there is a slight drop in classification accuracy. This can be due to certain details of the implementation of low-level runtime drivers.

\textbf{\texttt{BUFFER\_NUM}.} Results of varying the parameter \texttt{BUFFER\_NUM} is shown in Figure~\ref{FigBUFFERNUM}. %In Figure~\ref{FigBUFFERNUM}~(a) - (d), as longer traces are involved, the layering phenomenon is more significant. \texttt{BUFFER\_NUM} corresponds to the length of the trace. 
By increasing \texttt{BUFFER\_NUM}, a longer period of execution of the victim accelerators will be probed and the trace can include more information. However, surprisingly, from Figure~\ref{FigAccuracy}~(d), the classification accuracy does not change much. With \texttt{BUFFER\_NUM}$ =50$, our random forest classifier is able to reach over $90\%$. Other models have a similar trend of accuracy.

\textbf{Summary.}
In our experimental results, we show that for a relatively wide range of parameter choices, random forest model is able to achieve satisfying classification accuracy. This helps loose the constraints on attackers' benchmark accelerator implementation. Under \texttt{ACCESS\_NUM}$ =1000$, \texttt{REPEAT\_NUM}$ =5$, \texttt{BUFFER\_NUM}$ =100$, and \texttt{BUFFER\_SIZE}$ =4$Bytes our model is able to achieve the highest accuracy. However, in real world, under some other hardware or software settings (different FPGA models, communication link hardware, or a different heterogeneous computing software stack) these values may vary. To maximize attack performance, attackers are recommended to conduct some offline screening prior to launching the attack to obtain near-optimal parameters. This parameter search does not need to be accurate, since our most powerful model can achieve over $92\%$ accuracy performance under a relatively wide range of attack accelerator parameter choices in the selected accelerator set, which is sufficient for fingerprinting tasks. From the benchmark accelerator side, we conclude that:
\begin{enumerate}
    \item Our benchmark accelerator is able to capture the I/O patterns of each of the victim accelerators.
    \item Both \texttt{ACCESS\_NUM} and \texttt{REPEAT\_NUM} affect the granularity of measurement and can significantly influence the performance of classification models. There are optimal values for these two values, as shown in Figure~\ref{FigAccuracy}~(a) and Figure~\ref{FigAccuracy}~(b).
    \item Buffer related parameters \texttt{BUFFER\_SIZE} and \texttt{BUFFER\_NUM} have less influence on classification accuracy. However, the optimal parameter values still exist.
\end{enumerate}

%In the answers to RQ $2$, we show how different parameters in our benchmark accelerator implementation (\texttt{ACCESS\_NUM}, \texttt{REPEAT\_NUM}, \texttt{BUFFER\_NUM}, \texttt{BUFFER\_SIZE}) can affect the collected traces hence influencing the classifier performance. 

\section{Discussion}\label{SecDiscussion}
\subsection{Mitigation}
The intrinsic cause of the security vulnerability revealed in this paper is the different communication or I/O patterns of accelerators. The different access patterns of accelerators can serve as unique fingerprints of these accelerators. What our benchmark accelerator and host program do is to stress the communication link (i.e. PCIe) and obtain performance measurement trace results that contain information about these fingerprints. This information is further extracted by machine learning models and helps achieve high classification accuracy.

The mitigation to our proposed fingerprint attack can be done by enhancing the FPGA-host interface [e.g. FPGA interface manager (FIM) in Intel cloud FPGAs~\cite{devcloud}]. Instead of transmitting raw data, messages travelling through the communication link should pass another security layer for obfuscation. In this obfuscation layer, the communication pattern will be distorted, where random latency/burst will be inserted to make communication patterns unrecognizable. Policies targeting introducing such distortions with minimum performance overhead will be our future work.

From the host side, we can also modify the underlying platforms (OPAE~\cite{opae}, OpenCL~\cite{devcloudOpenCLGuide}, etc.). By changing how the driver handles data movement between host server and FPGA, communication pattern obfuscation can also be achieved.

% \subsection{Parameter Choice}
% In the answers to RQ $2$, we show how different parameters in our benchmark accelerator implementation (\texttt{ACCESS\_NUM}, \texttt{REPEAT\_NUM}, \texttt{BUFFER\_NUM}, \texttt{BUFFER\_SIZE}) can affect the collected traces hence influencing the classifier performance. In our experimental results, we show that for a relatively wide range of parameter choices, random forest model is able to achieve satisfying classification accuracy. This helps loose the constraints on attackers' benchmark accelerator implementation.

% In our experiments, under \texttt{ACCESS\_NUM}$ =1000$, \texttt{REPEAT\_NUM}$ =5$, \texttt{BUFFER\_NUM}$ =100$, and \texttt{BUFFER\_SIZE}$ =4$Bytes our model is able to achieve the highest accuracy. However, in real world, under some other hardware or software settings (different FPGA models, communication link hardware, or a different heterogeneous computing software stack) these values may vary. Attackers are advised to conduct some offline screening prior to launching the attack to obtain near-optimal parameters. This parameter search does not need to be accurate, since our most powerful model can achieve over $92\%$ accuracy performance under a relatively wide range of attack accelerator parameter choices, which is sufficient for fingerprinting tasks.

\subsection{Attack and Defence Suggestions}
\subsubsection{For Attackers} In our proposed attack, one prerequisite for attackers is to obtain servers and FPGAs that are identical to the servers and FPGAs used in FPGA clouds. In reality, instead of purchasing hardware and building up the system locally, it's better for attackers to use the cloud itself for data collection. By running data collection steps in the cloud multiple times and recording the underlying hardware and software platform, the attackers can eventually have a set of models that are able to cover heterogeneous hardware and software platforms in the cloud. Doing this step on the victim cloud is more realistic and economic, considering the high cost to set up required hardware and software environments locally.

\subsubsection{For Regular FPGA Cloud Users} The fingerprinting attack we propose relies on the intrinsic features of victim accelerators, and we make an assumption that attackers are aware of the target accelerator and can limit the range of accelerators running on the cloud. Therefore, to defend against the proposed fingerprinting attack, FPGA cloud users should be careful about using existing public intellectual property cores (IPs) since these IPs are possibly already in the attackers' database. To achieve this, these users can modify their accelerators and insert noisy I/O or computation operations (additional writes to an unimportant memory location, inserting additional computation between two I/O operations, etc.) to distort the performance traces the attacker may obtain. 

In the meantime, it is worth noting that exploiting this security vulnerability also relies on physically residing on the same FPGA where victim accelerator is running. The simplest way for users to avoid being attacked is to obtain ownership of the whole FPGA board as well as the hosting server. This may result in higher costs in deploying FPGA accelerators (since it requires users to pay more to cloud service providers), but it completely eliminates the threat of side-channel-related attacks induced by sharing FPGA resources with unknown users.

\subsubsection{For Cloud Service Providers} We suggest cloud service providers enhance their infrastructure interface as mentioned in Section~\ref{SecDiscussion}.1. Though this may add additional performance overhead, it can efficiently prevent the proposed attack.

Also, FPGA cloud service providers can consider improving their scheduling policy to scatter users' FPGA accelerators on the cloud. It can dramatically reduce the chance of victims' malicious accelerators co-locating with victims' accelerators and hence mitigating side-channel attacks or fingerprinting attacks that require attackers and victims to be placed together.

\subsection{Future Work}
Future work will be dedicated to developing mitigation technologies against this side-channel. We will come up with both hardware and software-based mitigation strategies. For hardware defence, we will consider deploying low-overhead noise injection circuits. For software defence, we will modify the underlying heterogeneous computing frameworks like OpenCL~\cite{munshi2009opencl} or OPAE~\cite{opae} to obfuscate the communication patterns of computation accelerators on board. Besides, cloud scheduler-level defence can be employed to securely schedule/migrate instances.

\section{Related Work}\label{SecRelatedWork}
\textbf{FPGA side-channel attacks.}
Several kinds of remote attacks targeting cloud FPGAs have been proposed recently. One major type is long-wire attack, where attackers utilize leakage in long wires to probe information transmitted inside the circuit. \cite{longwirebit} uses the delay difference of nearby wires to probe the signal being transmitted on the long wire, since logical $1$ and logical $0$ on long wires can lead to different delays of nearby wires. The authors use ROs to capture this difference and use collected information to recover the bits being transmitted on the target long-wire. \cite{sidechannelwithoutphysical} performs a similar attack and recovers the secret key of an AES crypt circuit. \cite{giechaskiel2019measuring} provides detailed tests of several RO designs and validate the efficiency of these variants of long-wire attacks. Defence mechanisms are proposed as well to mitigate long-wire attacks. Remote power side-channel attack is another type of FPGA side-channel attack. In~\cite{powersidechannel}, it is performed by programming an on-chip RO-based power monitor to reveal the secret key of a RSA crypto module. This paper also shows that by using the RO-based power monitor, it is possible to perform an FPGA-to-CPU attack on the same SoC. Power side-channel attacks have also been proven to be feasible in production environment~\cite{glamovcanin2020cloud}, where researchers retrieved AES key information from an AWS EC2 F1 FPGA instance.

Our attack is based on PCIe communication side-channels. There has been attack in FPGA cloud using PCIe contention. In~\cite{tian2021cloud}, the authors utilize the generation of PCIe contention to perform infrastructure cartography. They use PCIe stressors to generate PCIe contention and reveal information regarding cloud servers in AWS cloud. However, their attack targets multiple FPGAs and aim at revealing infrastructure information instead of revealing information about applications on the same FPGA. In~\cite{giechaskiel2021cross,giechaskiel2022cross}, the authors build a covert communication channel based on PCIe contention and consider information leakage in the PCIe contention side-channel. Similar to our work, PCIe traffic is monitored, and information like execution timing traces of victim applications can be obtained. The difference is that we consider multi-tenancy FPGAs (accelerators from multiple users residing on the same FPGA hardware), whereas they consider the scenario where accelerators from different users are distributed to multiple FPGA boards connected to the same server.

The most similar works we find in literature are~\cite{gobulukoglu2021classifying,drewes2023turn}. To the best of our knowledge, they are also the only works about multi-tenancy FPGA accelerator fingerprinting. In these papers, to achieve a similar goal, the authors propose to use power side-channel for fingerprinting co-located FPGA circuits. Their measurement targets lower-level side-channel leakage and they focus on classifying cryptographic cores, whereas our method is more coarse-grained and we focus on identifying general accelerator workloads.

Our proposed method is more closely related and will be beneficial to side-channel attacks in FPGA cloud, which rely on co-locating with target victims and information about co-located victim circuits. These attacks include attacks targeting cross-talk information leakage~\cite{giechaskiel2020information}, power analysis attacks~\cite{moini2021remote,provelengios2019characterizing} that collects power side-channel information using co-located malicious circuits and reveal secret information from collected data, fault attacks~\cite{alam2019ram} that actively induces faults like voltage drops to victim circuits, etc.

\section{Conclusion}\label{SecConclusion}
In this paper, we propose a novel attack targeting multi-tenancy FPGA clouds, where attackers can obtain knowledge about co-located accelerators. By implementing a PoC attack accelerator as well as its corresponding host program, we test accelerators from several application scenarios like signal processing, numerical simulation acceleration, etc. Our results show that communication links like PCIe can serve as a new source of side-channel and can be exploited by fingerprinting attacks targeting co-located FPGA accelerators. Our proposed attack method will be beneficial for cloud FPGA side-channel attacks, since successfully recognizing target co-located victims is a prerequisite and can significantly reduce the costs of attacks. As far as we know, this is the first work targeting fingerprinting co-located FPGA accelerators using communication side-channels. Future work will be dedicated to security-enhanced FPGA interface development and another version of this research under an open-world setting.

\balance
\bibliographystyle{ACM-Reference-Format}
\bibliography{sample-sigconf}


\end{document}